\documentclass{article}
\usepackage{amsmath,amsthm,amsfonts,amscd,eucal}

\numberwithin{equation}{section}

\def\bn{{\mathbb N}}

\def\r{\rho}

\def\itm#1{\item{$(#1)$}}

\DeclareMathOperator{\Var}{Var}
\DeclareMathOperator{\Cov}{Cov}

\newtheorem{theorem}{Theorem}[section]
\newtheorem{proposition}[theorem]{Proposition}
\newtheorem{lemma}[theorem]{Lemma}

\newtheorem{corollary}[theorem]{Corollary}
\theoremstyle{definition}
\newtheorem{definition}[theorem]{Definition}
\newtheorem{example}[theorem]{Example}
\newtheorem{remark}[theorem]{Remark}

\makeindex

\begin{document}

\title{
{Uncertainty Principle and Quantum Fisher Information - II}
}

\author{
Paolo Gibilisco\footnote{Dipartimento SEFEMEQ, Facolt\`a di
Economia, Universit\`a di Roma ``Tor Vergata", Via Columbia 2, 00133
Rome, Italy.  Email:  gibilisco@volterra.uniroma2.it --
URL: http://www.economia.uniroma2.it/sefemeq/professori/gibilisco},
Daniele Imparato\footnote{Dipartimento di Matematica, Politecnico di Torino,
Corso Duca degli Abruzzi 24, 10129 Turin, Italy.
Email: daniele.imparato@polito.it}
\
and
Tommaso Isola\footnote{Dipartimento di Matematica,
Universit\`a di Roma ``Tor Vergata",
Via della Ricerca Scientifica, 00133 Rome, Italy.
Email: isola@mat.uniroma2.it
URL: http://www.mat.uniroma2.it/$\sim$isola}
}

\maketitle

\begin{abstract}

Heisenberg and Schr{\"o}dinger uncertainty principles give lower bounds
for the product of variances ${\rm Var}_{\rho}(A)\cdot{\rm
Var}_{\rho}(B)$ if the observables $A,B$ are not compatible, namely if
the commutator $[A,B]$ is not zero.

In this paper we prove an uncertainty principle in Schr{\"o}dinger
form where the bound for the product of variances ${\rm
Var}_{\rho}(A)\cdot{\rm Var}_{\rho}(B)$ depends on the area spanned by
the commutators $i[\rho,A]$ and $i[\rho,B]$ with respect to an arbitrary
quantum version of the Fisher information.

\end{abstract}

\smallskip
\noindent 2000 {\sl Mathematics Subject Classification.} Primary 62B10, 94A17; Secondary 46L30, 46L60.

\noindent {\sl Key words and phrases.} Uncertainty principle, means, monotone metrics,
quantum Fisher information, Wigner-Yanase-Dyson information.

\section{Introduction}
Let $X,Y$ be random variables on a probability space $(\Omega, {\cal
G}, p)$ and consider the covariance ${\rm Cov}_p(X,Y):={\rm
E}_p(XY)-{\rm E}_p(X){\rm E}_p(Y)$ and the variance ${\rm
Var}_{p}(X):={\rm Cov}_p(X,X)$.  The best one can get from
Cauchy-Schwartz inequality is the following inequality
$$
{\rm Var}_{p}(X)\cdot{\rm Var}_{p}(Y)-|{\rm Cov}_{p}(X,Y)|^2
\geq
0. \eqno{(1.1)}
$$
The uncertainty principle is one of the most striking consequences
of non-commutativity in Quantum Mechanics and is a key point in
which quantum probability differs from classical probability.  We
shall limit our discussion to the matrix case.  If $\rho$ is a state
(i.e. density matrix), $A,B$ observables (i.e. self-adjoint
matrices), set ${\rm Cov}_{\rho}(A,B):={\rm Tr}(\rho A B)-{\rm
Tr}(\rho A)\cdot{\rm Tr}(\rho B)$.  Define also the symmetrized
covariance as ${\rm Cov}^s_{\rho}(A,B):= \frac12 [{\rm
Cov}_{\rho}(A,B)+{\rm Cov}_{\rho}(B,A)]$ and the variance as ${\rm
Var}(A):={\rm Cov}_{\rho}(A,A)={\rm Cov}^s_{\rho}(A,A)$.  Again from
Cauchy-Schwartz inequality one gets the following inequality
$$
{\rm Var}_{\rho}(A)\cdot{\rm Var}_{\rho}(B)-|{\rm Cov}^s_{\rho}(A,B)
|^2 \geq \frac{1}{4}\vert {\rm Tr}(\rho[A,B])\vert^2, \eqno{(1.2)}
$$
which is known as the Schr{\"o}dinger uncertainty principle.  By
omitting the covariance part, one gets the Heisenberg uncertainty
principle (see \cite{Heisenberg:1927}\cite{Schrodinger:1930}).
Inequality (1.2) states that the condition $[A,B]\not=0$ ($i.e.$
$A,B$ are not compatible) gives a limitation to the simultaneous
``smallness" of both ${\rm Var}_{\rho}(A)$ and ${\rm Var}_{\rho}(B)$
and this has very important consequences in Quantum Mechanics.

But non-commutativity can enter also from another side.  One may
naturally ask if there are similar bounds for the product ${\rm
Var}_{\rho}(A)\cdot {\rm Var}_{\rho}(B)$ due to the fact that the
observables $A,B$ do not commute with the state $\rho$.  Indeed this
is the case, and our main result will provide such a bound in terms
of an ``area" spanned by the commutators $i[\rho,A]$ and
$i[\rho,B]$.

To state our result we have to introduce the notion of quantum
Fisher information.  Let us denote by ${\cal F}_{op}$ the class of
normalized symmetric operator monotone functions on $(0,+\infty)$.
It is a by now classical result of Petz that to each function $f \in
{\cal F}_{op}$ one can associate a Riemannian metric $\langle A, B
\rangle_{\rho, f}$ on the state manifold that is monotone and
therefore a quantum version of the Fisher information (see
\cite{Chentsov:1982} \cite{Petz:1996}).  For example, the functions
$$
f_\beta(x):= \beta (1- \beta) \frac{(x-1)^2}{(x^{\beta}-1)
(x^{1-\beta}-1)}\qquad \qquad x>0, \quad\beta \in (0,1/2],
$$
are associated to a quantum Fisher information that is related to
the well known Wigner-Yanase-Dyson ($WYD$) skew information.
Indeed, for the $WYD$-information of parameter $\beta$ one has

$$
-\frac{1}{2}{\rm Tr}([\rho^{\beta},A]\cdot[\rho^{1-\beta},A]) =
\frac{\beta(1-\beta)}{2} \langle i[\rho, A],i[\rho, A] \rangle_{\rho,
f_{\beta}}.
$$
We denote by $\hbox{Area}^f_{\rho}(u,v)$ the area spanned by the
tangent vectors $u,v$ with respect to the Riemannian monotone metric
associated to $f$ (at the point $\rho$). For $f \in {\cal F}_{op}$
let us define $f(0):= \displaystyle\lim_{x \to 0}f(x)$; $f$ is said
regular
 if $f(0)>0$ (see \cite{Hansen:2006b}). The subset of regular elements of ${\cal F}_{op}$ is denoted by ${\cal F}_{op}^{\, r}$.

The
goal of the present paper is to prove the following inequality
$$
{\rm Var}_{\rho}(A)\cdot{\rm Var}_{\rho}(B)-|{\rm Cov}^s_{\rho}(A,B)
|^2 \geq \frac{1}{4}\left(f(0) \cdot {\rm
Area}_{\rho}^f(i[\rho,A],i[\rho,B])\right)^2 \qquad \forall f \in {\cal F}_{op}^{\, r}.  \eqno{(1.3)}
$$

Note that inequality (1.3) holds trivially in the non-regular case.
Our result has been inspired by particular cases of the above
theorem that have been proved recently.  Luo and Z. Zhang
\cite{LuoZZhang:2004} conjectured the inequality (1.3) for the
Wigner-Yanase metric, namely for the function $f_{1/2}$.  This
conjecture was proved shortly after by Luo himself and Q. Zhang
\cite{LuoQZhang:2004}.  The case of Wigner-Yanase-Dyson metric
(namely the metric associated to $f_\beta$ for $\beta \in (0, 1/2)$)
was proved independently by Kosaki \cite{Kosaki:2005} and by Yanagi
{\sl et al.} \cite{YanagiFuruichiKuriyama:2005}.  In our paper
\cite{GibiliscoIsola:2006b} we emphasized the geometric aspects of
the question and we succeeded to formulate (1.3) for a general
quantum Fisher information.

It is worth to emphasize the dynamical meaning of the inequality
(1.3). Indeed, each positive (self-adjoint) operator $H$ determines
a time evolution of a state $\rho$ according to the formula
$\rho_H(t):=e^{-iHt} \rho e^{iHt}$.  If $[\rho,H]=0$, then there is
no evolution.  Therefore, we may say that the bound given by (1.3)
appears for those pairs of observables that are ``dynamically
incompatible", that is, for pairs $H,K$ such that the associated
evolutions $\rho_H(t),\rho_K(t)$ are different and non-trivial (this
is equivalent to the linear independence of $[\rho,H]$ and
$[\rho,K]$).

As a by-product of the work needed to prove our main result, we
derive also other two inequalities interesting {\sl per se}. Indeed,
a crucial ingredient in the proof of (1.3) is the following formula
$$
\tilde{f}(x):=\frac{1}{2}\left[ (x+1)-(x-1)^2 \frac{f(0)}{f(x)}
\right],
$$
that associates to any element $f \in {\cal F}_{op}$ another element
${\tilde f} \in {\cal F}_{op}$. Let $A_0:=A-{\rm Tr}(\rho A)I$ and
denote by $L_{\rho}, R_{\rho}$ respectively the left and right
multiplication by $\rho$. Let $m_f$ be the mean associated to $f$
(see Section \ref{m} below) and define
$$
{\cal C}^f_{\rho}(A_0):={\rm Tr}(m_{f}(L_{\rho},R_{\rho})(A_0) \cdot
A_0), \qquad \qquad I_{\rho}^f(A):={\rm Var}_{\rho}(A)-{\cal
C}^{\tilde f}_{\rho}(A_0).
$$
Hansen introduced $I^f_{\rho}(A)$ in the paper \cite{Hansen:2006b} with a different approach. We shall call it ``metric adjusted skew information" or
``$f$-information" to stress the dependence on the function $f$.
One may consider $I_{\rho}^f(A)$ as a generalization of Wigner-Yanase-Dyson information.
Let $f_{RLD}(x)=\frac{2x}{x+1}$; we prove the following inequality
$$
{\rm Var}_{\rho}(A)\cdot{\rm Var}_{\rho}(B) \geq [I_{\rho}^f(A)+{\cal C}^{f_{RLD}}_{\rho}(A_0)]\cdot[I_{\rho}^f(B)+{\cal C}^{f_{RLD}}_{\rho}(B_0)]
\qquad \forall f \in {\cal F}_{op}^{\, r}. \eqno{(1.4)}
$$
Moreover, we also prove that, if $f \in {\cal F}_{op}$,
$$
{\rm Var}_{\rho}(A)\cdot{\rm Var}_{\rho}(B) \geq {\cal C}^f_{\rho}(A_0){\cal C}^f_{\rho}(B_0)+\frac{1}{4}\vert {\rm Tr}(\rho[A,B])\vert^2 \quad
\Longleftrightarrow \quad  \quad f(x)\leq \sqrt{x}. \eqno{(1.5)}
$$
Inequality (1.4) is a refinement of an inequality proved by Luo, for
the Wigner-Yanase metric, and by Hansen, in the general case (see
\cite{Luo:2003}, \cite{Hansen:2006b}). Inequality (1.5) for the
function $\sqrt{x}$ is  due to Park and, independently, to Luo (see
\cite{Park:2005} \cite{Luo:2005b}). Here we simply prove the
optimality of their bound in ${\cal F}_{op}$.

The plan of the paper is the following.

Sections \ref{HS}, \ref{KA}, \ref{P} contain preliminary notions.
In Section \ref{HS} we recall the standard Heisenberg and
Schr{\"o}dinger uncertainty principles.  In Section \ref{KA} we give
the fundamental definitions and theorems for number and operator
means.  In Section \ref{P} we review the classification theorem for
the quantum Fisher informations.

Sections \ref{h}, \ref{m}, \ref{corr}, \ref{e} contain the core of
the paper.  In Section \ref{h} we show that to any operator monotone
function $f \in {\cal F}_{op}$ one may associate another element
${\tilde f} \in {\cal F}_{op}$ by formula (5.1); we study the
properties of the mean $m_{\tilde f}$ and of an associated function
$H_f$, discussing, in particular, how they behave as functions of
$f$. In Section \ref{m} we prove the main result, namely the
inequality (1.3). Furthermore, we study the right side of the above
inequality as a function of $f$ and relate it to quantum evolution
of states, as said before.  In Section \ref{corr} we introduce the
$f$-correlation (a kind of generalized Wigner-Yanase-Dyson
correlation) and the $f$-information and discuss their relation with
the quantum Fisher information associated to $f\in {\cal F}_{op}$.
In this way, we are able to show how the inequality (1.3)
generalizes the previously known results. Moreover, we prove that by
choosing the SLD (Bures-Uhlmann) metric the lower bound given in
(1.3) is optimal and strictly greater than the previously known
optimal bound (given by the Wigner-Yanase metric). In Section
\ref{e} we prove a necessary and sufficient condition to get the
equality in (1.3).

In Section \ref{another} we prove the inequality (1.4).  In Section
\ref{counter} we produce counterexamples to prove the logical
independence of the uncertainty principles studied in this paper -
that is, inequalities (1.3) and (1.4) - from the standard
Heisenberg-Schr{\"o}dinger uncertainty principles.  In Section
\ref{p} we discuss what happens for not faithful and pure states,
also at the light of the notion of radial extension for quantum
Fisher information. In Section \ref{Park} we show the optimality of
an improvement of Heisenberg uncertainty principle recently proposed
by Park and Luo, namely we prove the inequality (1.5).

\section{Heisenberg and Schr{\"o}dinger Uncertainty Principles} \label{HS}

Let $M_n:=M_n(\mathbb{C})$ (resp. $M_{n,sa}:=M_n(\mathbb{C})_{sa}$)
be the set of all $n \times n$ complex matrices (resp.  all $n
\times n$ self-adjoint matrices).  We shall denote general matrices
by $X,Y,...$ while letters $A,B,...$ will be used for self-adjoint
matrices.  The Hilbert-Schmidt scalar product is denoted by $\langle
A,B \rangle={\rm Tr}(A^*B)$.  The adjoint of a matrix $X$ is denoted
by $X^{\dag}$ while the adjoint of a superoperator $T:(M_n,\langle
\cdot,\cdot \rangle) \to ( M_n ,\langle \cdot,\cdot \rangle)$ is
denoted by $T^*$. Let ${\cal D}_n$ be the set of strictly positive
elements of $M_n$ and ${\cal D}_n^1 \subset {\cal D}_n$ be the set
of strictly positive density matrices; namely,
$$
{\cal D}_n^1=\{\rho \in M_n \vert {\rm Tr} \rho=1, \, \rho>0 \}.
$$
From now on, we shall treat the case of faithful states, namely
$\rho>0$. We shall consider the general case $\rho \geq 0$ at the
end of the paper, in Section \ref{p}, where we shall also discuss in
detail what happens for pure states.

\begin{definition}
Suppose that $\rho \in {\cal D}_n^1$ is fixed.  Define $X_0:=X-{\rm
Tr}(\rho X) I$.
\end{definition}

\begin{definition}
For $A,B \in M_{n,sa}$ and $\rho \in {\cal D}_n^1$ define covariance
and variance as $$\begin{array}{rcl} {\rm Cov}_{\rho}(A,B)&:=&{\rm
Tr}(\rho A B)-{\rm Tr}(\rho A)\cdot{\rm Tr}(\rho B)= {\rm Tr}(\rho
A_0 B_0),\\[12pt] {\rm Var}_{\rho}(A)&:=&{\rm Tr}(\rho A^2)-{\rm Tr}(\rho
A)^2 = {\rm Tr}(\rho A^2_0).\end{array}
$$
\end{definition}

\begin{proposition} \label{cov}
$$\begin{array}{rcl}
2{\rm Re}\{ {\rm Cov}_{\rho}(A,B) \}& =& {\rm Cov}_{\rho}(A,B)+{\rm
Cov}_{\rho}(B,A)={\rm Tr}(\rho \{A_0, B_0\}),\\[12pt] 2i{\rm Im}\{ {\rm
Cov}_{\rho}(A,B) \}& = &{\rm Cov}_{\rho}(A,B)-{\rm
Cov}_{\rho}(B,A)={\rm Tr}(\rho [A, B]),\end{array}
$$
where, for any $X,Y\in M_{n}$, $[X,Y] := XY-YX$, $\{X,Y\} := XY+YX$.
\end{proposition}

We define the symmetrized covariance as ${\rm Cov}^s_{\rho}(A,B):=
\frac{1}{2} [{\rm Cov}_{\rho}(A,B)+{\rm Cov}_{\rho}(B,A)]={\rm Re}\{
{\rm Cov}_{\rho}(A,B) \}$.  The Cauchy-Schwartz inequality implies
$$
|{\rm Cov}_{\rho}(A,B)|^2 \leq {\rm Var}_{\rho}(A){\rm Var}_{\rho}(B).
$$
From this one gets the Schr{\"o}dinger and Heisenberg uncertainty
principles which are stated in the following theorem.

\begin{theorem} (see \cite{Schrodinger:1930})

For $A,B \in M_{n,sa}$ and $\rho \in {\cal
D}_n^1$ one has
$$
{\rm Var}_{\rho}(A){\rm Var}_{\rho}(B)-|{\rm Cov}^s_{\rho}(A,B)|^2
\geq
\frac{1}{4}\vert {\rm Tr}(\rho[A,B])\vert^2,
$$
that implies
$$
{\rm Var}_{\rho}(A){\rm Var}_{\rho}(B)
\geq
\frac{1}{4}\vert {\rm Tr}(\rho[A,B])\vert^2.
$$
\end{theorem}

\section{Means for positive numbers and matrices} \label{KA}

For this Section we refer to the exposition contained in
\cite{PetzTemesi:2005}.

\begin{definition} \label{numean}

Let ${\mathbb R}^+:= (0,+\infty)$.  A {\sl mean} for pairs of
positive numbers is a function $m:{\mathbb R}^+ \times {\mathbb R}^+
\to{\mathbb R}^+$ such that

\itm{i} $m(x,x)=x$,

\itm{ii} $m(x,y)=m(y,x)$,

\itm{iii} $x <y  \quad \Longrightarrow \quad x<m(x,y)<y$,

\itm{iv} $x<x', \quad y<y' \quad \Longrightarrow \quad m(x,y)<m(x',y')$,

\itm{v} $m$ is continuous,

\itm{vi} for $t>0$ one has $m(tx,ty)=t \cdot m(x,y)$.

We denote by $\displaystyle {\cal M}_{nu}$ the set of means.
\end{definition}

\begin{definition} \label{numon}
${\cal F}_{nu}$ is the class of functions $f: {\mathbb R}^+
\to{\mathbb R}^+$ such that

\itm{i} $f(1)=1$,

\itm{ii} $tf(t^{-1})=f(t)$,

\itm{iii} $t \in (0,1) \Longrightarrow f(t) \in (0,1)$,

\itm{iv} $t \in (1,\infty) \Longrightarrow f(t) \in (1,\infty)$,

\itm{v} $f$ is continuous,

\itm{vi} $f$ is monotone increasing.

\end{definition}

\begin{proposition}
There is bijection between ${\cal M}_{nu}$ and ${\cal F}_{nu}$ given
by the formulas
$$
m_f(x,y):=yf(xy^{-1}), \qquad\qquad f_m(t):=m(1,t).
$$
\end{proposition}

\begin{remark}
$$
f \leq  g \Longleftrightarrow m_f \leq m_g.
$$
\end{remark}

Here below we report the Kubo-Ando theory of matrix means (see
\cite{KuboAndo:1979/80}) as exposed in \cite{PetzTemesi:2005}. In
the sequel, for any pairs of matrices $A$, $B$, we shall write $A<B$
whenever $B-A$ is positive semidefinite.

\begin{definition}

Recall that ${\cal D}_n:= \{A \in M_n({\mathbb C}) | A>0 \}$.  A {\sl
mean} for pairs of positive matrices is a function
$m(\cdot,\cdot):{\cal D}_n \times {\cal D}_n \to {\cal D}_n$ such that conditions $(i)--(v)$ of Definition \ref{numean} hold (with the matrix partial order defined above) and the
{\sl transformer inequality}

$Cm(A,B)C^* \leq m(CAC^*,CBC^*)$, $\forall C$,

\noindent
replaces $(vi)$. We denote by $\displaystyle {\cal M}_{op}$ the set of matrix means.

\end{definition}

\begin{example}
The  arithmetic, geometric and harmonic (matrix) means are given respectively by
$$\begin{array}{rcl}
A \nabla B&:=&\frac{1}{2}(A+B),\\[12pt]A\#
B&:=&A^{\frac{1}{2}}(A^{-\frac{1}{2}} B
A^{-\frac{1}{2}})^{\frac{1}{2}}A^{\frac{1}{2}},
\\[12pt]
A{\rm !}B&:=&2(A^{-1}+B^{-1})^{-1}. \end{array}$$
\end{example}

Let us recall that a function $f:(0,\infty)\to \mathbb{R}$ is said
{\it operator monotone} if, for any $n\in \bn$, any $A$, $B\in M_n$
such that $0\leq A\leq B$, the inequalities $0\leq f(A)\leq f(B)$
hold.  An operator monotone function is said {\it symmetric} if
$f(x)=xf(x^{-1})$ and {\it normalized} if $f(1)=1$.

\begin{definition}
${\cal F}_{op}$ is the class of operator monotone functions $f: {\mathbb R}^+
\to{\mathbb R}^+$ such that conditions $(i)--(v)$ of Definition \ref{numon} hold (with the matrix partial order defined above).

\end{definition}

Note that the above definition is redundant (see for example
\cite{Bhatia:1996}); however, it well emphasizes the similarity with
the number case.  Indeed, one has the following result.

\begin{proposition}

${\cal F}_{op}$ is the class of functions $f: {\mathbb R}^+
\to{\mathbb R}^+$ such that

\itm{i'} $f(1)=1$,

\itm{ii'} $tf(t^{-1})=f(t)$,

\itm{iii'} $f$ is operator monotone increasing.

Equivalently, $f \in {\cal F}_{op}$ iff $f$ is a normalized,
symmetric, operator monotone function.

\end{proposition}

The fundamental result, due to Kubo and Ando, is the following.

\begin{theorem}
There is bijection between ${\cal M}_{op}$ and ${\cal F}_{op}$ given by
the formula

$$
m_f(A,B):= A^{\frac{1}{2}}f(A^{-\frac{1}{2}} B
A^{-\frac{1}{2}})A^{\frac{1}{2}}.
$$
\end{theorem}

When $A$ and $B$ commute, we have that

$$
m_f(A,B):= A \cdot f( B  A^{-1}).
$$

\begin{theorem}
Among matrix means, arithmetic is the largest while harmonic is the
smallest.
\end{theorem}
\begin{proof}
See Theorem 4.5 in \cite{KuboAndo:1979/80}.
\end{proof}

\begin{corollary} \label{basic}
For any $f \in {\cal F}_{op}$ and for any $x,y>0$ one has
$$
\frac{2x}{1+x}\leq f(x) \leq \frac{1+x}{2},
$$
$$
\frac{2xy}{x+y}\leq m_f(x,y) \leq \frac{x+y}{2}.
$$
\end{corollary}

\section{Quantum Fisher Informations} \label{P}

In what follows, given a differential manifold ${\cal N}$, we denote
by $T_{\rho} \cal N$ the tangent space to $\cal N$ at the point
$\rho \in {\cal N}$. In the commutative case a Markov morphism is a
stochastic map $T: \mathbb{R}^n \to \mathbb{R}^m$. Let
 $$
\displaystyle{ {\cal P}_n := \{ \rho \in \mathbb{R}^n \vert \rho_i >
0 \}, \qquad {\cal P}_n^1 :=\{ \rho \in \displaystyle{\cal P}_n\vert
\sum \rho_i=1 \}}.$$ The natural representation for the tangent
space is given by
$$
T_{\rho}{\cal P}^1_n =\{v \in \mathbb{R}^n| \sum_i v_i=0 \}.
$$
 In this case a monotone metric is defined as a family of Riemannian
 metrics $g=\{g^n\}$ on $\{{\cal P}_n^1\}$, $n \in \mathbb{N}$, such
 that
 $$
 g^m_{T(\rho)}(TX,TX) \leq g^n_{\rho}(X,X)
 $$
 holds for every Markov morphism $T:\mathbb{R}^n \to \mathbb{R}^m$,
 for every $\rho \in {\cal P}_n^1$ and for every $X \in T_\rho {\cal
 P}_n^1$.

The Fisher information is the Riemannian metric on ${\cal P}_n^1$ defined as
$$
\langle u,v \rangle_{\rho, F}:=\sum_i \frac{u_iv_i}{\rho_i} \qquad \qquad u,v \in T_\rho {\cal P}_n^1.
$$

\begin{theorem} (see \cite{Chentsov:1982})

    There exists a unique monotone metric on ${\cal P}_n^1$ (up to scalars) given by
    the Fisher information.
\end{theorem}

In the noncommutative case a Markov morphism is a completely
positive and trace preserving operator $T: M_n \to M_m$. Recall that
there exists a natural identification
 of $T_{\rho}{\cal D}^1_n$ with the space of self-adjoint traceless
 matrices, namely for any $\rho \in {\cal D}^1_n $
$$
T_{\rho}{\cal D}^1_n =\{A \in M_{n, sa}| \hbox{Tr}(A)=0 \}.
$$

 In perfect analogy with the commutative case, a monotone metric in
 the noncommutative case is a family of Riemannian metrics $g=\{g^n\}$
 on $\{{\cal D}^1_n\}$, $n \in \mathbb{N}$, such that
 $$
 g^m_{T(\rho)}(TX,TX) \leq g^n_{\rho}(X,X)
 $$
 holds for every Markov morphism $T:M_n \to M_m$, for every $\rho \in
 {\cal D}^1_n$ and for every $X \in T_\rho {\cal D}^1_n$.
Monotone metrics are usually normalized in such a way that
$[A,\rho]=0$ implies $g_{f,\rho} (A,A)={\rm Tr}({\rho}^{-1}A^2)$.

 To a normalized symmetric operator monotone function $f\in {\cal
 F}_{op}$ one associates the so-called CM (Chentsov--Morozowa)
 function
 $$
 c_f(x,y):=\frac{1}{yf(xy^{-1})}=m_f(x,y)^{-1}\qquad\hbox{for}\qquad
 x,y>0.
 $$
Define $L_{\rho}(A):= \rho A$, and $R_{\rho}(A):= A\rho$; observe
 that they are self-adjoint operators on $M_{n,sa}$.  Since $L_{\rho}$
 and $R_{\rho}$ commute we may define
 $c_f(L_{\rho},R_{\rho})=m_f(L_{\rho},R_{\rho})^{-1}$.  Since $m_f$ is
 a matrix mean one gets the following result.

\begin{proposition} (see \cite{Petz:1996}) \label{sa}

    $m_f(L_{\rho},R_{\rho})$ and $c_f(L_{\rho},R_{\rho})$ are positive
    and therefore self-adjoint.
\end{proposition}
Now we can state the fundamental theorems about noncommutative monotone metrics.

\begin{theorem} (see \cite{Petz:1996})

    There exists a bijective correspondence between monotone metrics
    on ${\cal D}^1_n$ and normalized symmetric operator monotone
    functions $f\in {\cal F}_{op}$.  This correspondence is given by
    the formula
    $$
   \langle A,B \rangle_{\rho,f}:={\rm Tr}(A\cdot
    c_f(L_\rho,R_\rho)(B))={\rm Tr}(A\cdot
    m_f(L_{\rho},R_{\rho})^{-1}(B)).
    $$
\end{theorem}
We set $||A||^2_{\rho,f} := \langle A,A \rangle_{\rho,f}$.
Because of the above theorems we shall use the terms ``Monotone
Metrics" and ``Quantum Fisher Informations" (shortly QFI) with the
same meaning.

For a symmetric operator monotone function define
$f(0):=\displaystyle\lim_{x \to 0}f(x) = \displaystyle\lim_{x \to
+\infty} \frac{f(x)}{x}$.  Of course, $f(0) \geq 0 $.  
The condition $f(0)\not=0$ is relevant because it is a necessary and
sufficient condition for the existence of the so-called radial
extension of a monotone metric to pure states (see
\cite{PetzSudar:1996}\cite{PetzSudar:1999} or Section \ref{p} below).
Following \cite{Hansen:2006b} we say that a function $f \in {\cal
F}_{op}$ is {\sl regular} iff $f(0) \not= 0$.  The corresponding
operator mean, CM function, associated QFI, etc.  are said regular
too. The class of regular (resp. non-regular) functions $f \in {\cal F}_{op}$ is denoted by ${\cal F}_{op}^{\, r}$ (resp. ${\cal F}_{op}^{\, n}$).

As proved by Lesniewski and Ruskai each quantum Fisher information is
the Hessian of a suitable relative entropy (see
\cite{LesniewskiRuskai:1999}).

\section{The function ${\tilde f}$ and the properties of the
associated mean} \label{h}

In \cite{Hansen:2006b} it has been proved the following result.

\begin{proposition} (Proposition 3.4 in \cite{Hansen:2006b})

If $f\in {\cal F}_{op}$ is regular, define the representing function as
$$
d_f(x,y):= \frac{(x+y)}{f(0)}-(x-y)^2c_f(x,y), \qquad \qquad x,y>0.
$$

Then, the function $d_f$ is positive and operator concave.
\end{proposition}

\begin{definition}
For $f \in {\cal F}_{op}$ and $x>0$ set
$$
\tilde{f}(x):=\frac{1}{2}\left[ (x+1)-(x-1)^2 \frac{f(0)}{f(x)}
\right].  \eqno{(5.1)}
$$
\end{definition}

\begin{proposition}
$$
f \in {\cal F}_{op} \qquad \Longrightarrow \qquad {\tilde f} \in {\cal
F}_{op}.
$$
\end{proposition}
\begin{proof}
Easy calculations show that $\tilde f$ is normalized and symmetric.
To prove that $f$ is operator monotone note that:

$(a)$ if $f$ is not regular then ${\tilde f}(x)=\frac{1}{2}(1+x)$ and the
conclusion follows;

$(b)$ if $f$ is regular then ${\tilde f}(x)=\frac{f(0)}{2}d(x,1)$.
Since $d$ is positive and operator concave so is ${\tilde f}$.  We get
the conclusion because operator concavity is equivalent to operator
monotonicity (see \cite{HansenPedersen:1982}).
\end{proof}

\begin{remark}
    Note that $f$ regular  $\Longrightarrow$ ${\tilde f}$ not regular.
\end{remark}

\smallskip

Following the terminology of Section \ref{KA} we associate to
$\tilde f$ both a number and an operator mean by the formulas
$$\begin{array}{rcl}
m_{\tilde f}(x,y)&:= &y \cdot {\tilde f}( x  y^{-1}),\\  m_{\tilde
f}(A,B)&:= &A^{\frac{1}{2}}{\tilde f}(A^{-\frac{1}{2}} B
A^{-\frac{1}{2}})A^{\frac{1}{2}}. \end{array}$$

\begin{remark}
    Observe that $\displaystyle m_{\tilde f}(x,y) = \frac{x+y}{2}
    -\frac{f(0)}{2} \frac{(x-y)^{2}}{yf(\frac{x}{y})}$.
\end{remark}

From Corollary \ref{basic} one obtains this result.

\begin{corollary} \label{max}

    For any $f \in {\cal F}_{op}$ and for any $x,y>0$ one has
$$
\frac{2x}{1+x}\leq \tilde{f}(x) \leq \frac{1+x}{2},
$$
$$
\frac{2xy}{x+y}\leq m_{\tilde f}(x,y) \leq \frac{x+y}{2}.
$$
\end{corollary}

Moreover we have the following result, whose proof is elementary.

\begin{proposition} \label{tilde}

    For every $x>0$ and $f,g \in {\cal F}_{op}$
$$
\tilde{f}(x) \leq \tilde{g}(x) \quad \Longleftrightarrow \quad
\frac{f(0)}{f(x)} \geq \frac{g(0)}{g(x)}.
$$
\end{proposition}

We synthetize some results in the following Table.

\centerline{\textsc{Table I}}

\bigskip

\begin{tabular}{|c|c|c|c|c|c|}
\hline
&&&&&\\
QFI  & $f$ & $m_f$ & $f(0)$& $\tilde{f}$ & $m_{\tilde f}$  \\ 
&&&&&\\
\hline 
 
&&&&&\\
$RLD$ & $\frac{2x}{x+1}$ & $\frac{2}{\frac{1}{x}+\frac{1}{y}}$ & $0$& $\frac{1+x}{2}$ & $\frac{x+y}{2}$  \\
&&&&&\\
\hline

&&&&&\\
$WYD(\beta)$
& $\frac{ \beta (1- \beta) (x-1)^2}{(x^{\beta}-1) (x^{1-\beta}-1)}$ 
&$\frac{ \beta (1- \beta) (x-y)^2}{(x^{\beta}-y^{\beta}) (x^{1-\beta}-y^{1-\beta})}$ 
 & $0$ &
$\frac{1+x}{2}$
& $\frac{x+y}{2}$  \\

$ \beta \in (-1,0) $ &&&&&\\
&&&&&\\
\hline

&&&&&\\
$BKM$ & $\frac{x-1}{\log x}$ & $\frac{x-y}{\log x - \log y}$ & $0$& $\frac{1+x}{2}$ & $\frac{x+y}{2}$ \\
&&&&&\\
\hline

&&&&&\\
$WYD(\beta)$
& $\frac{ \beta (1- \beta) (x-1)^2}{(x^{\beta}-1) (x^{1-\beta}-1)}$ 
&$\frac{ \beta (1- \beta) (x-y)^2}{(x^{\beta}-y^{\beta}) (x^{1-\beta}-y^{1-\beta})}$ & $\beta(1-\beta)$& 
$\frac{x^{\beta}+x^{1-\beta}}{2}$
& $\frac{x^{\beta}y^{1-\beta}+x^{1-\beta}y^{\beta}}{2}$ \\
$ \beta \in (0,\frac{1}{2})$ &&&&&\\
&&&&&\\
\hline

&&&&&\\
$WY$ & $\left( \frac{1+\sqrt{x}}{2}\right)^2$  &  $\left( \frac{\sqrt{x}+\sqrt{y}}{2}\right)^2$ & $\frac{1}{4}$ & $\sqrt{x}$ &  $\sqrt{xy}$  \\ 
&&&&&\\
\hline

&&&&&\\
$SLD$ & $\frac{1+x}{2}$ & $\frac{x+y}{2}$ & $\frac{1}{2}$& $\frac{2x}{x+1}$ & $\frac{2}{\frac{1}{x}+\frac{1}{y}}$  \\
&&&&&\\ 
\hline  

\end{tabular}

\smallskip

\smallskip

{\scriptsize In the above table we have, for some quantum Fisher informations: the name,
the function $f$, the mean $m_f$, the value of $f$ at 0, the function
$\tilde{f}$ and the mean $m_{\tilde{f}}$.}

\bigskip

\begin{example}
Let $x>0$ and $\beta \in (0,\frac{1}{2})$. If we set 
$$
f_{SLD}(x):=\frac{1+x}{2}, \quad f_{WY}(x):=\left(
\frac{1+\sqrt{x}}{2} \right)^2, \quad f_{\beta}(x):=\beta (1- \beta)
\frac{(x-1)^2}{(x^{\beta}-1) (x^{1-\beta}-1)}, \quad
f_{RLD}(x):=\frac{2x}{1+x}.
$$
One has (see the above table)
$$
\tilde{f}_{SLD}(x)=\frac{2x}{1+x}, \quad
\tilde{f}_{WY}(x)=\sqrt{x}, \quad
\tilde{f}_{\beta}(x):=\frac{x^{\beta}+x^{1-\beta}}{2}, \quad
\tilde{f}_{RLD}(x):=\frac{1+x}{2}.
$$

Note that if  $x>0$ is fixed the function $ \beta\in
    (0,\frac{1}{2}) \mapsto x^{\beta}+x^{1-\beta}\in {\mathbb R}^+$ is
    decreasing. This implies
$$
\tilde{f}_{SLD} \leq \tilde{f}_{WY}\leq \tilde{f}_{\beta}\leq
\tilde{f}_{RLD},
$$
 and therefore
$$
m_{\tilde{f}_{SLD}} \leq m_{\tilde{f}_{WY}} \leq
m_{\tilde{f}_{\beta}} \leq m_{\tilde{f}_{RLD}},
$$
that is a refined arithmetic-geometric-harmonic inequality
$$
\frac{2xy}{x+y} \leq \sqrt{xy} \leq
\frac{1}{2}(x^{\beta}y^{1-\beta}+x^{1-\beta}y^{\beta})\leq
\frac{1+x}{2} \qquad \qquad x,y>0, \quad \beta \in (0, 1/2).
$$
\end{example}

\begin{remark}
\end{remark}
The metrics
associated with the functions $f_{\beta}$ are equivalent to the
metrics induced by noncommutative $\alpha$-divergences, where
$\beta=\frac{1-\alpha} {2}$ (see \cite{HasegawaPetz:1997}).  They
are very important in information geometry and are related to
Wigner-Yanase-Dyson information (see for example
\cite{GibiliscoIsola:2003}\cite{GibiliscoIsola:2004}
\cite{GibiliscoIsola:2005}).
Defining $\ell_{\gamma}(x):=((1+x^{\gamma})/2)^{\frac{1}{\gamma}}$ for $\gamma \in [1/2,1]$ one has $\ell_{\gamma} \in {\cal F}_{op}$.
The two parametric families $f_{\beta}, \ell_{\gamma}$ give us a
continuum of operator monotone functions from the smallest function
$\frac{2x}{x+1}$ to the largest function
$\frac{1+x}{2}$.  Further examples of this kind of ``bridges"
can be found in \cite{Hansen:2006a} \cite{Hansen:2006b}. Note that also $g_0(x):= \sqrt{x}$ is an element of ${\cal F}_{op}$.

\bigskip

In the sequel we need to study the following function.

\begin{definition}
For any $f \in {\cal F}_{op}$ set
$$
H_{ f}(x,y,w,z):=[(x+y)-m_{\tilde f}(x,y)]m_{\tilde
f}(w,z)+[(w+z)-m_{\tilde f}(w,z)]m_{\tilde f}(x,y) \qquad \qquad
x,y,w,z > 0.
$$
\end{definition}

\begin{proposition} \label{monarea}

For any $f,g \in {\cal F}_{op}$

$$\begin{array}{rcl}
\tilde{f} &\leq & \tilde{g}
\\ &
\Downarrow &\\ H_{f}(x,y,w,z) &\leq &H_g(x,y,w,z) \qquad \qquad
\forall x,y,w,z >0. \end{array}$$
\end{proposition}

\begin{proof}

Since
$$
(x+y)-m_{\tilde f}(x,y)=(x+y)-\frac{x+y}{2}+ \frac{(x-y)^2}{2y} \cdot
\frac{f(0)}{f(\frac{x}{y}) } =\frac{x+y}{2}+
\frac{(x-y)^2}{2y} \cdot \frac{f(0)}{f(\frac{x}{y}) }
$$
we have
\begin{equation}\begin{array}{rl}
    H_{f}(x,y,w,z) & :=\displaystyle[(x+y)-m_{\tilde f}(x,y)]m_{\tilde
f}(w,z)+[(w+z)-m_{\tilde f}(w,z)]m_{\tilde f}(x,y)\\[12pt]
& =\displaystyle\left(\frac{x+y}{2}+ \frac{(x-y)^2}{2y} \cdot
\frac{f(0)}{f( \frac{x}{y}) }\right) \cdot \left( \frac{w+z}{2}-
\frac{(w-z)^2}{2z} \cdot \frac{f(0)}{f(\frac{w}{z}) }
\right) \\[12pt]
& \displaystyle\quad +\left( \frac{w+z}{2}+ \frac{(w-z)^2}{2z} \cdot
\frac{f(0)}{f( \frac{w}{z}) } \right) \cdot \left(\frac{x+y}{2}-
\frac{(x-y)^2}{2y} \cdot \frac{f(0)}{f( \frac{x}{y})
}\right) \\[12pt]
& =\displaystyle\frac{1}{2} \left[ (x+y)(w+z) - \left(
\frac{(x-y)^2}{y} \frac{(w-z)^2}{z} \right) \left( \frac{f(0)}{f(
\frac{x}{y})} \cdot \frac{f(0)}{f(\frac{w}{z}) } \right)
\right].\label{H}
\end{array}\end{equation}

Since, from Proposition \ref{tilde},
$$
\tilde{f} \leq \tilde{g} \Rightarrow \frac{f(0)}{f(t)} \geq
\frac{g(0)}{g(t)} > 0 \qquad \qquad \forall t >0,
$$ we obtain
$$
H_{ f}(x,y,w,z) \leq H_g(x,y,w,z) \qquad \qquad \forall x,y,w,z >0
$$
by elementary computations.
\end{proof}

Note that for $f$ non-regular one has
$$
H_{ f}(x,y,w,z)=\frac{1}{2}(x+y)(w+z).
$$
On the other hand, for the function $f_{SLD}=\frac{1}{2}(1+x)$ one
has from (\ref{H})
$$
H_{SLD}(x,y,w,z) = \frac{1}{2} \left[ (x+y)(w+z)-\frac{1}{4} \left(
\frac{(x-y)^2(w-z)^2}{\frac{x+y}{2} \cdot \frac{w+z}{2}}\right)\right]
= 2 \frac{xy(w^2+z^2)+wz(x^2+y^2)}{(x+y)(w+z)}.
$$
Therefore, we have the following bounds.

\begin{corollary} \label{mon2}

For any $f \in {\cal F}_{op}$
$$
0< 2\left[ \frac{xy(w^2+z^2)+wz(x^2+y^2)}{(x+y)(w+z)}\right]\leq H_{
f}(x,y,w,z) \leq \frac{1}{2}(x+y)(w+z)\qquad \qquad \forall x,y,w,z
>0.
$$
\end{corollary}

\begin{remark}  \label{sldwy}

    Note that for every $x>0$
$$
\frac{f_{SLD}(0)}{f_{SLD}(x)}=
\frac{\frac{1}{2}}{\frac{1+x}{2}}=\frac{1}{1+x}>\frac{1}{1+x+2\sqrt{x}}
=\frac{1}{(1+\sqrt{x})^2}=\frac{\frac{1}{4}}{\frac{(1+\sqrt{x})^2}{4}}
= \frac{f_{WY}(0)}{f_{WY}(x)},
$$
so that for every $x,y,w,z>0$
$$
H_{ SLD}(x,y,w,z)<H_{ WY}(x,y,w,z).
$$
\end{remark}

\section{The main result} \label{m}

\begin{proposition} \label{delta}

Given  $f \in {\cal F}_{op}$ , let $\Delta:=m_{\tilde
f}(L_{\rho},R_{\rho})$. Recall that $B_0:=B-{\rm Tr}(\rho B)$. One
has

\itm{i} ${\rm Tr}(B_0 \cdot \Delta(I))=0$,

\itm{ii} ${\rm Tr}(I\cdot  \Delta(B_0))=0$,

\itm{iii} ${\rm Tr}(\Delta(I))=1$.
\end{proposition}

\begin{proof}

    $(i)$ Since $(L_{\rho}-R_{\rho})(I)=0$ and ${\rm Tr}(\rho
    B_0)={\rm Tr}(\rho B)-{\rm Tr}(\rho B)=0$ we have

\begin{align*}
    \langle B_0, m_{\tilde f}(L_{\rho},R_{\rho}) (I)\rangle &={\rm
    Tr}(B_0 m_{\tilde f}(L_{\rho},R_{\rho})(I)) \\
    &=\frac{1}{2}{\rm Tr}(B_0
    (L_{\rho}+R_{\rho})(I))-\frac{1}{2}f(0){\rm Tr}(B_0
    {c}_f(L_{\rho},R_{\rho})(L_{\rho}-R_{\rho})^2(I)) \\
    &=\frac{1}{2}{\rm Tr}(B_0 \rho+\rho B_0)={\rm Tr}(\rho B_0) \\
    &=0.
\end{align*}

$(ii)$ It is a simple consequence of $(i)$ and of Proposition
\ref{sa}. Indeed,
$$
\langle I, m_{\tilde f}(L_{\rho},R_{\rho}) (B_0)\rangle= \langle
m_{\tilde f}(L_{\rho},R_{\rho})(I), B_0\rangle=0.
$$

$(iii)$
\begin{align*}
    {\rm Tr}(\Delta(I)) &={\rm Tr}(m_{\tilde f}(L_{\rho},R_{\rho})(I))
    \\
    &=\frac{1}{2}{\rm Tr}((L_{\rho}+R_{\rho})(I))-\frac{1}{2}f(0){\rm
    Tr}({c}_f(L_{\rho},R_{\rho})(L_{\rho}-R_{\rho})^2(I)) \\
    &={\rm Tr}(\rho) \\
    &=1.
\end{align*}

\end{proof}

\begin{proposition} \label{boh}
$$
f(0) \cdot \langle i[\rho,A], i[\rho,B]\rangle_{\rho, f}={\rm Tr}(\rho
AB)+ {\rm Tr}(\rho BA) - 2 {\rm Tr}( A \cdot \Delta(B)).
$$
\end{proposition}

\begin{proof}
Let us introduce the shorthand notation
$$
\hat{c}_f(x,y):=(x-y)^2c_f(x,y),
$$
so that by definition
$$
f(0) \cdot {\hat c}_f(x,y)= (x+y) - 2 m_{\tilde f}(x,y).
$$

Therefore, we have

\begin{align*}
    f(0) \cdot \langle i[\rho,A], i[\rho,B]\rangle_{\rho, f} &=f(0)
    \cdot {\rm Tr}\bigl( (i[\rho,A]) \cdot c_f(L_{\rho},R_{\rho})
    (i[\rho,B]) \bigr) \\
    &=f(0) \cdot \langle (i[\rho,A]), c_f(L_{\rho},R_{\rho})
    (i[\rho,B]) \rangle \\
    &=f(0) \cdot \langle i(L_{\rho}-R_{\rho})(A),
    c_f(L_{\rho},R_{\rho})\circ (i(L_{\rho}-R_{\rho}))(B) \rangle \\
    &=f(0) \cdot \langle A, (i(L_{\rho}-R_{\rho}))^* \circ
    c_f(L_{\rho},R_{\rho})\circ (i(L_{\rho}-R_{\rho}))(B) \rangle \\
    &=f(0) \cdot \langle A, -i(L_{\rho}-R_{\rho}) \circ
    c_f(L_{\rho},R_{\rho})\circ (i(L_{\rho}-R_{\rho}))(B) \rangle \\
    &=f(0) \cdot \langle A, {\hat c}_f(L_{\rho},R_{\rho})(B) \rangle
    \\
    &= f(0) \cdot {\rm Tr}(A \cdot {\hat c}_f(L_{\rho},R_{\rho})(B))
    \\
    &= {\rm Tr}(A \cdot (f(0)\cdot {\hat c}_f(L_{\rho},R_{\rho}))(B))
    \\
    &={\rm Tr}(A \cdot (L_{\rho}+R_{\rho} - 2m_{\tilde
    f}(L_{\rho},R_{\rho}))(B)) \\
    &={\rm Tr}(\rho AB)+ {\rm Tr}(\rho BA) - 2 {\rm Tr}( A \cdot
    m_{\tilde f}(L_{\rho},R_{\rho})(B))).
\end{align*}
\end{proof}

\begin{proposition} \label{nonso}
$$
f(0) \cdot \langle i[\rho,A], i[\rho,B]\rangle_{\rho, f}=2\bigl( {\rm
Re} \{ {\rm Cov}_{\rho} ( A,B)\} -{\rm Tr}(\Delta( A_0)B_0)
\bigr).
$$
\end{proposition}
\begin{proof}
We have that
\begin{align*}
    f(0) \cdot \langle i[\rho,A], i[\rho,B]\rangle_{\rho, f} &={\rm
    Tr}(\rho AB)+ {\rm Tr}(\rho BA) - 2 {\rm Tr}( A \cdot \Delta(B))
    \\
    &= {\rm Cov}_{\rho} \left( {A,B} \right)+{\rm Cov}_{\rho} \left(
    {B,A} \right)+2{\rm Tr}(\rho A)\cdot{\rm Tr}(\rho B)- 2 {\rm Tr}(
    A \cdot \Delta(B))\\
    &=2{\rm Re} \{ {\rm Cov}_{\rho} ( A,B) \} + 2 \bigl( {\rm Tr}(\rho
    A)\cdot{\rm Tr}(\rho B)- {\rm Tr}( A \cdot \Delta(B)) \bigr);
\end{align*}

moreover, because of Proposition \ref{delta},
\begin{align*}
    {\rm Tr}(\rho A ){\rm Tr}(\rho B)-{\rm Tr}(\Delta( A) B) &={\rm
    Tr}(\rho A ){\rm Tr}(\rho B)-{\rm Tr}(\Delta( A_0+{\rm Tr}(\rho
    A)I) (B_0+{\rm Tr}(\rho B)I)) \\
    &={\rm Tr}(\rho A ){\rm Tr}(\rho B) -\bigl[ {\rm Tr}(\Delta(
    A_0)B_0)+{\rm Tr}(\rho A){\rm Tr}(\Delta(I) B_0) \\
    & \qquad + {\rm Tr}(\Delta( A_0)I){\rm Tr}(\rho B)+{\rm Tr}(\rho
    A){\rm Tr}(\rho B){\rm Tr}(\Delta( I)I) \bigr] \\
    &={\rm Tr}(\rho A ){\rm Tr}(\rho B)-{\rm Tr}(\Delta( A_0)B_0)-{\rm
    Tr}(\rho A){\rm Tr}(\rho B) \\
    &=-{\rm Tr}(\Delta( A_0)B_0).
\end{align*}

Therefore, the conclusion follows.

\end{proof}

We recall some consequences of the spectral theorem we need in the
sequel.  Let $\rho$ be a state, $\lambda_i$ its eigenvalues and
$E_i$ the associated eigenprojectors.  The spectral decompositions
of $L_{\rho}$ and $R_{\rho}$ are the following
$$
L_{\rho}= \sum_{i} \lambda_i L_{E_i} \qquad \qquad
R_{\rho}= \sum_{i} \lambda_i R_{E_i}.
$$
Therefore, from the spectral theorem for commuting selfadjoint
operators we get the following result.

\begin{corollary} \label{coro:spectral}

    Let $\rho$ be a state, $\lambda_i$ its eigenvalues and $E_i$ the
    projectors of the associated eigenspaces.  If $s: [0,+\infty)
    \times [0,+\infty) \to \mathbb{R}$ is a continuous function then
$$
s(L_{\rho},R_{\rho})=\sum_{i,j} s(\lambda _i, \lambda _j )L_{E_i}R_{E_j}.
$$
\end{corollary}

Let $V$ be a finite dimensional real vector space with a scalar
product $g(\cdot,\cdot)$.  We define, for $v,w \in V$,
$$
\hbox{Area}^g(v,w):=\sqrt{g(v,v) \cdot g(w,w)- \vert g(v,w) \vert^2}.
$$
In the Euclidean plane $\hbox{Area}^g(v,w)$ is the area of the
parallelogram spanned by $v$ and $w$. If we are dealing with a
$\rho$ point-depending Riemannian metric, we write
$\hbox{Area}^g_{\rho}$. If $f \in {\cal F}_{op}$ we denote by ${\rm
Area}_{\rho}^f$ the area functional associated to the monotone
metric $\langle \cdot, \cdot \rangle_{\rho,f}$.

We are now ready for the main results.

\begin{theorem} \label{main}
For any $f,g \in {\cal F}_{op}$

\itm{i}
$$
{\rm Var}_\rho ( A ){\rm Var}_\rho ( B ) - |{\rm Cov}^{s}_{\rho}
(A,B)|^2 \geq \left( \frac{f(0)}{2 } \cdot {\rm
Area}^{f}_{\rho}(i[\rho,A],i[\rho,B]) \right)^2,
$$

\itm{ii}
$$
\tilde{g} \geq \tilde{f}\qquad \Longrightarrow \qquad
\frac{g(0)}{2 } \cdot {\rm
Area}^{g}_{\rho}(i[\rho,A],i[\rho,B]) \leq \frac{f(0)}{2 } \cdot {\rm
Area}^{f}_{\rho}(i[\rho,A],i[\rho,B]).
$$
\end{theorem}

\begin{proof}
Fix $A,B\in M_{n,sa}$. Let us introduce, for the sake of brevity,
$$
F(f):= \left( {\rm Var}_\rho ( A ){\rm Var}_\rho ( B ) - |{\rm
Cov}^{s}_{\rho}(A,B)|^2 \right) -\left( \frac{f(0)}{2 } \cdot {\rm
Area}^{f}_{\rho}(i[\rho,A],i[\rho,B]) \right)^2.
$$
Then, we have to show that $F(f)\geq0$, and $\tilde{g} \geq
\tilde{f}\Longrightarrow F(g) \geq F(f)$.

Let $\left\{\varphi_i\right\}$ be a complete orthonormal base
composed of eigenvectors of $\rho$ and $\{ {\lambda}_i \}$ the
corresponding eigenvalues. Set $a_{ij} \equiv \langle {A_0}
{\varphi}_i |{\varphi}_j \rangle $ and $ b_{ij} \equiv \langle B_0
\varphi_i | {\varphi_j } \rangle $. Note that $a_{ij} \not=
A_{ij}:=$ the $i,j$ entry of $A$.

Then we calculate
\begin{align*}
    \Var_{\rho}(A) &= {\rm Tr} ( \rho A_0^2) = \frac{1}{2}\sum_{i,j} (
    \lambda _i + \lambda_j ) a_{ij} a_{ji} \\
    \Var_{\rho} ( B ) &= {\rm Tr} (\rho B_0^2 ) =
    \frac{1}{2}\sum_{i,j} (\lambda _i + \lambda _j )b_{ij} b_{ji} \\
    {\rm Cov}^{s}_{\rho} (A,B)&={\rm Re}\{ \Cov_{\rho}(A,B) \} = {\rm
    Re} \{{\rm Tr} (\rho A_0 B_0) \}= \frac{1}{2}
    \sum_{i,j}({\lambda}_i+{\lambda}_j) {\rm Re} \{a_{ij}b_{ji} \} \\
    \frac{f(0)}{2}||i[\rho,A]||^2_{\rho,f}
    &= \Var_{\rho} ( A ) - {\rm Tr} ( A_0 m_{\tilde f}(L_{\rho},
    R_{\rho}) A_0) = \frac{1}{2}\sum_{i,j} ( \lambda _i + \lambda_j )
    a_{ij} a_{ji} -\sum_{i,j} m_{\tilde f}(\lambda_i,\lambda_j)
    a_{ij}a_{ji} \\
    \frac{f(0)}{2}||i[\rho,B]||^2_{\rho,f}
      &=\frac{1}{2}\sum_{i,j} (\lambda _i + \lambda _j )b_{ij} b_{ji} -
    \sum_{i,j} m_{\tilde f}(\lambda_i,\lambda_j)b_{ij} b_{ji} \\
    \frac{f(0)}{2}\langle i[\rho,A], i[\rho,B]\rangle_{\rho,f}
     &= {\rm Re} \{ {\rm Cov}_{\rho} ( {A,B} ) \} - {\rm Re} \{ {\rm
    Tr}(m_{\tilde f}(L_{\rho}, R_{\rho}) (A_0) \cdot B_0) \} \\
    &= \frac{1}{2} \sum_{i,j}({\lambda}_i+{\lambda}_j) {\rm Re} \{
    a_{ij} b_{ji} \} - \sum_{i,j} m_{\tilde f}(\lambda_i,\lambda_j)
    {\rm Re} \{ {a_{ij} b_{ji} } \} .
\end{align*}
Set
\begin{align*}
    \xi &:= \hbox{Var}_\rho \left( A \right) \hbox{Var}_\rho \left( B
    \right) - \frac{f(0)^2}{4}||i[\rho,A]||^2_{\rho,f} \cdot
    ||i[\rho,B]||^2_{\rho,f} \\
    &=\frac{1}{2}\sum_{i,j,k,l} \left\{ ( \lambda _i + \lambda _j
    )m_{\tilde f}(\lambda_k,\lambda_l) + ( \lambda _k + \lambda _l
    )m_{\tilde f}(\lambda_i,\lambda_j)- 2m_{\tilde
    f}(\lambda_i,\lambda_j) m_{\tilde f}(\lambda_k,\lambda_l) \right\}
    a_{ij} a_{ji} b_{kl} b_{lk} \\
    &=\frac{1}{4}\sum_{i,j,k,l} \left\{ ( \lambda _i + \lambda _j
    )m_{\tilde f}(\lambda_k,\lambda_l) + ( \lambda _k + \lambda _l
    )m_{\tilde f}(\lambda_i,\lambda_j) - 2m_{\tilde
    f}(\lambda_i,\lambda_j) m_{\tilde f}(\lambda_k,\lambda_l) \right\}
    \{ a_{ij} a_{ji} b_{kl} b_{lk}+ a_{kl}a_{lk}b_{ij}b_{ji} \}, \\
    \eta &:= |{\rm Cov}^{s}_{\rho}(A,B)|^2 -\frac{f(0)^2}{4} | \langle
    i[\rho,A], i[\rho,B]\rangle^2_{\rho,f}|^2 \\
    &= \frac{1}{2}\sum_{i,j,k,l} \left\{ ( \lambda _i + \lambda _j)
    m_{\tilde f}(\lambda_k,\lambda_l) + ( \lambda _k + \lambda _l )
    m_{\tilde f}(\lambda_i,\lambda_j) - 2m_{\tilde
    f}(\lambda_i,\lambda_j) m_{\tilde f}(\lambda_k,\lambda_l) \right\}
    {\rm Re} \{ a_{ij} b_{ji} \} {\rm Re} \{ a_{kl} b_{lk} \}, \\
    K_{i,j,k,l}&:= K_{i,j,k,l}(\rho,A,B):= |a_{ij} |^2 |b_{kl}|^2 +
    |a_{kl} |^2 |b_{ij}|^2 - 2{\rm Re} \{ a_{ij} b_{ji} \} {\rm Re} \{
    a_{kl} b_{lk} \} .
\end{align*}

Since
$$
|a_{ij} |^2 |b_{kl}|^2 + |a_{kl} |^2 |b_{ij}|^2 \ge 2\left| {a_{ij}
b_{ji} } \right|\left| {a_{kl} b_{lk} } \right| \ge 2\left|
{{\mathop{\rm Re}\nolimits} \left\{ {a_{ij} b_{ji} }
\right\}{\mathop{\rm Re}\nolimits} \left\{ {a_{kl} b_{lk} } \right\}}
\right|,
$$
we have that $K_{i,j,k,l} \geq 0$.  Note that $K_{i,j,k,l}$ does not
depend on $f$.

Then
\[
\begin{array}{rcl}
    F(f)= \xi - \eta &=&\dfrac{1}{4}\displaystyle\sum_{i,j,k,l} \left\{ ( \lambda _i
    + \lambda _j )m_{\tilde f}(\lambda_k,\lambda_l) + ( \lambda _k +
    \lambda _l )m_{\tilde f}(\lambda_i,\lambda_j) - 2m_{\tilde
    f}(\lambda_i,\lambda_j) m_{\tilde f}(\lambda_k,\lambda_l) \right\}
    \cdot \\[12pt]
    &&\cdot \left\{ |a_{ij} |^2 |b_{kl}|^2 + |a_{kl} |^2 |b_{ij}|^2 -
    2{\rm Re} \{ a_{ij} b_{ji} \} {\rm Re} \{ a_{kl} b_{lk} \}
    \right\} \\[12pt]
    &=&\dfrac{1}{4}\displaystyle\sum_{i,j,k,l}H_{
    f}(\lambda_i,\lambda_j,\lambda_k,\lambda_l) \cdot K_{i,j,k,l}.
\end{array}
\]

Because of Proposition \ref{monarea} and Corollary \ref{mon2} one has
that
$$
\tilde{f} \leq \tilde{g} \quad \Longrightarrow \quad 0 \leq H_{
f}(\lambda_i,\lambda_j,\lambda_k,\lambda_l) \leq
H_g(\lambda_i,\lambda_j,\lambda_k,\lambda_l)
$$
and therefore
$$
\tilde{f} \leq \tilde{g} \quad \Longrightarrow \quad 0 \leq F(f) \leq
F(g)
$$
and we get the thesis.
\end{proof}

The standard Schr\"odinger uncertainty principle reads as
$$
{\rm Area}^{{\rm Cov}^s}_{\rho}(A,B) \geq \frac{1}{2} |{\rm Tr}(\rho
[A,B])|,
$$
while the main result of the present paper can be expressed as
$$
{\rm Area}^{{\rm Cov}^s}_{\rho}(A,B) \geq \frac{f(0)}{2} \cdot {\rm
Area}^{f}_{\rho}(i[\rho,A],i[\rho,B]).
$$

\begin{corollary}
    For any $f \in {\cal F}_{op}$, $A,B\in M_{n,sa}$, one has
$$
\frac{f_{SLD}(0)}{2} \cdot {\rm
Area}^{f_{SLD}}_{\rho}(i[\rho,A],i[\rho,B]) \geq \frac{f(0)}{2} \cdot
{\rm Area}^{f}_{\rho}(i[\rho,A],i[\rho,B]).
$$
\end{corollary}
\begin{proof}
Immediate consequence of Corollary \ref{max}.
\end{proof}

\begin{remark}
Setting
$$
N_{\rho}^f(A,B):={\rm Area}^{{\rm Cov}^s}_{\rho}(A,B) - \frac{f(0)}{2 }
\cdot {\rm Area}^{f}_{\rho}(i[\rho,A],i[\rho,B]) \geq 0,
$$
we may strengthen the main result to
$$
{\rm Area}^{{\rm Cov}^s}_{\rho}(A,B) \geq \frac{f(0)}{2 } \cdot {\rm
Area}^{f}_{\rho}(i[\rho,A],i[\rho,B])+N_{\rho}^{SLD}(A,B).
$$
\end{remark}

The above geometric considerations take a particularly interesting
form when considering the dynamics of quantum states.  Suppose we have
a positive (self-adjoint) operator $H$ determining a quantum
evolution.  The state $\rho$ evolves according to the formula
$$
\rho_H(t):=e^{-itH} \rho e^{itH}.
$$
We say that $\rho_H(t)$ is the time evolution of $\rho=\rho_H(0)$
determined by $H$.  For the evolution $\rho_H(t)$ this is equivalent
to satisfy the quantum analogue of Liouville theorem in classical
statistical mechanics, namely the Landau-von Neumann equation.

\begin{definition}

Let $\r(t)$ be a curve in ${\cal D}_n^1$ and let $H \in M_{n,sa}$.
We say that $\r(t)$ satisfies the Landau-von Neumann equation w.r.t.
$H$ if

$$
\dot{\rho}(t)=\frac{{\rm d}}{{\rm d} t} \r(t)= i[\r(t),H].
$$
Satisfying the Landau-von Neumann equation is equivalent to
$\rho(t)=\rho_H(t)=e^{-itH} \rho e^{itH}$.
\end{definition}

From Theorem \ref{main} we get the following inequality.

\begin{proposition}
Let $\rho>0$ be a state and $H,K \in M_{n,sa}$.  Suppose that
$\rho=\rho_H(0)=\rho_K(0)$.  Then, for any $f \in {\cal F}_{op}$, one
has
$$
{\rm Area}^{{\rm Cov}^s}_{\rho}(H,K) \geq \frac{f(0)}{2 } \cdot {\rm
Area}^{f}_{\rho}(\dot{\rho}_H(0),\dot{\rho}_K(0)).
$$
\end{proposition}

Therefore, as we said in the Introduction, the bound on the right side
of our inequality appears when the evolutions $\rho_H(t),\rho_K(t)$ are
different and not trivial.

\section{The $f$-correlation associated to quantum Fisher
informations} \label{corr}

Mainly to confront our result with previous results we introduce the
notions of {\sl $f$-correlation} and {\sl $f$-information}.

\begin{definition}
$$\begin{array}{rcl}
{\cal C}^f_{\rho}(A,B)& = &{\cal C}^f_{\rho}(B,A) :={\rm Tr}(m_{
f}(L_{\rho},R_{\rho})(A) \cdot B)
\\[12pt]
{\cal C}^f_{\rho}(A)&:=&{\cal C}^f_{\rho}(A,A). \end{array}$$
\end{definition}

\begin{definition}
For $A,B \in M_{n,sa}$, $\rho \in {\cal D}_n^1$ and $f \in {\cal
F}_{op}$, the {\sl metric adjusted correlation} (or {\sl $f$-correlation}) and the {\sl metric adjusted skew information} (or {\sl $f$-information}) are
defined as
$$\begin{array}{rcl}
{\rm Corr}_{\rho}^{f}(A,B)&:= &{\rm Tr}(\rho AB)-{\cal C}^{\tilde
f}_{\rho}(A,B)={\rm Tr}(\rho AB)-{\rm Tr}(m_{\tilde
f}(L_{\rho},R_{\rho})(A) \cdot B),
\\[12pt]
I_{\rho}^f(A)&:=& {\rm Corr}_{\rho}^{f}(A,A). \end{array}$$
\end{definition}

The definition of ${\rm Corr}_{\rho}^{f}(A,B)$ appeared
in \cite{Hansen:2006b} in a different form.
For the $f$-correlation there is an analogue of Proposition \ref{cov}
for covariance.

\begin{lemma} \label{Corr}

For any $A,B \in M_{n,sa}$, $\rho \in {\cal D}_n^1$ and $f \in {\cal
F}_{op}$ one has

$$\begin{array}{rcl}
2{\rm Re} \{ {\rm Corr}_{\rho}^{f} ( A,B) \} &=&{\rm
Corr}_{\rho}^{f}(A,B)+{\rm Corr}_{\rho}^{f}(B,A)= f(0) \cdot \langle
i[\rho,A], i[\rho,B]\rangle_{\rho, f}
\\[12pt]
2i{\rm Im} \{ {\rm Corr}_{\rho}^{f} ( A,B) \}& =&{\rm
Corr}_{\rho}^{f}(A,B)-{\rm Corr}_{\rho}^{f}(B,A)={\rm Tr}(\rho
[A,B]). \end{array}$$
\end{lemma}
\begin{proof}
We have that
$$
{\rm Corr}_{\rho}^{f}(A,B)-{\rm Corr}_{\rho}^{f}(B,A)= {\rm Tr}(\rho
[A,B]),
$$
which is purely imaginary.

This implies
$$
{\rm Re} \{ {\rm Corr}_{\rho}^{f} ( A,B) \} = {\rm Re} \{ {\rm
Corr}_{\rho}^{f} ( B,A) \},
$$
so that
$$
2{\rm Re} \{ {\rm Corr}_{\rho}^{f} ( A,B) \} ={\rm
Corr}_{\rho}^{f}(A,B)+{\rm Corr}_{\rho}^{f}(B,A),
$$
$$
2i{\rm Im} \{ {\rm Corr}_{\rho}^{f} ( A,B) \} ={\rm
Corr}_{\rho}^{f}(A,B)-{\rm Corr}_{\rho}^{f}(B,A).
$$
Since
$$
{\rm Corr}_{\rho}^{f}(A,B)+{\rm Corr}_{\rho}^{f}(B,A)={\rm Tr}(\rho
AB)+ {\rm Tr}(\rho BA) - 2 {\rm Tr}( A \cdot \Delta(B)),
$$
the conclusion follows from Proposition \ref{boh}.
\end{proof}

\begin{corollary}\label{positivity}
$$
I_{\rho}^f(A)= {\rm Corr}_{\rho}^{f}(A,A)= \frac{f(0)}{2} \cdot
\langle i[\rho,A], i[\rho,A]\rangle_{\rho, f} =\frac{f(0)}{2} \cdot ||
i[\rho,A]||^2_{\rho, f}.
$$
\end{corollary}

\begin{remark}
    If
$$
f_\beta(x):= \beta (1- \beta) \frac{(x-1)^2}{(x^{\beta}-1)
(x^{1-\beta}-1)}\qquad \qquad \beta \in \Bigl(0, \frac{1}{2}\Bigr],
$$
then
$$
I_{\rho}^{f_{\beta}}(A)= \frac{f_{\beta}(0)}{2}\hbox{Tr}(i[\rho,A]
c_{f_{\beta}}(L_{\rho},R_{\rho}) i[\rho,A])= -\frac{1}{2}{\rm
Tr}([\rho^{\beta},A]\cdot[\rho^{1-\beta},A]),
$$
so $I_{\rho}^{f_{\beta}}(A)$ coincides with the Wigner-Yanase-Dyson
skew information.
\end{remark}

Let us reformulate the main result in terms of $f$-correlation.

\begin{proposition}
    For any $f \in {\cal F}_{op}$   one has
$$
\left( \frac{f(0)}{2 } \cdot {\rm
Area}^{f}_{\rho}(i[\rho,A],i[\rho,B]) \right)^2= I_{\rho}^{f} \left( A
\right)I_{\rho}^{f} \left( B \right) - \left| {{\mathop{\rm
Re}\nolimits} \left\{ {{\rm Corr}_{\rho}^{f} \left( {A,B} \right)}
\right\}} \right|^2 .
$$
\end{proposition}
\begin{proof}
\begin{align*}
    \left(\frac{f(0)}{2 } \cdot {\rm
    Area}^{f}_{\rho}(i[\rho,A],i[\rho,B])\right)^2 &=\frac{f(0)^2}{4 }
    \left(\langle i[\rho,A], i[\rho,A]\rangle_{\rho, f} \cdot \langle
    i[\rho,B], i[\rho,B]\rangle_{\rho, f} - \langle i[\rho,A],
    i[\rho,B]\rangle^2_{\rho, f} \right) \\
    &= \left(\frac{f(0)}{2 }\cdot || i[\rho,A]||^2_{\rho, f}
    \right)\cdot \left( \frac{f(0)}{2 }\cdot || i[\rho,B]||^2_{\rho,
    f} \right)- \left(\frac{f(0)}{2}\langle i[\rho,A],
    i[\rho,B]\rangle_{\rho, f} \right)^2 \\
    &= I_{\rho}^{f} \left( A \right)I_{\rho}^{f} \left( B \right) -
    \left| {{\mathop{\rm Re}\nolimits} \left\{ {{\rm Corr}_{\rho}^{f}
    \left( {A,B} \right)} \right\}} \right|^2 .
\end{align*}
\end{proof}

Therefore, our main result states that
$$
{\rm Var}_\rho ( A ){\rm Var}_\rho ( B ) - |{\rm Re}\{{\rm
Cov}_{\rho}(A,B)\}|^2 \geq I_{\rho}^{f} \left( A \right)I_{\rho}^{f}
\left( B \right) - \left| {{\mathop{\rm Re}\nolimits} \left\{ {{\rm
Corr}_{\rho}^{f} \left( {A,B} \right)} \right\}} \right|^2.
$$

Recall that we  introduced, for fixed $\rho, A,B$, the functional
$$\begin{array}{rcl}
F(f) &=& {\rm Var}_\rho ( A ){\rm Var}_\rho ( B ) - |{\rm
Cov}^{s}_{\rho}(A,B)|^2  -\left( \dfrac{f(0)}{2 } \cdot {\rm
Area}^{f}_{\rho}(i[\rho,A],i[\rho,B]) \right)^2\\[12pt]
&=& {\rm Var}_\rho ( A ){\rm Var}_\rho ( B ) - |{\rm Re}\{{\rm
Cov}_{\rho}(A,B)\}|^2 -I_{\rho}^{f} \left( A \right)I_{\rho}^{f}
\left( B \right) + \left| {{\mathop{\rm Re}\nolimits} \left\{ {{\rm
Corr}_{\rho}^{f} \left( {A,B} \right)} \right\}} \right|^2.
\end{array}$$

As the main result, we proved that, for any $f,g \in {\cal F}_{op}$,
$ F(f) \geq 0$ and $\tilde{f} \leq \tilde{g} \Longrightarrow F(f)
\leq F(g)$.

\begin{corollary} \label{biggest}
Suppose $\rho, A, B$ are fixed. Then the function of $\beta$ given
by
$$
F(\beta):=F(f_{\beta})
$$
is decreasing on $(0,\frac{1}{2}]$ and $F(1/2)\geq 0$; therefore
$F(\beta)\geq 0$.
\end{corollary}
\begin{proof}
Given $x>0$, the function $\beta \mapsto {\tilde
f}_{\beta}(x)=\frac{1}{2}(x^{\beta}+x^{1-\beta})$ is decreasing in
$(0,\frac{1}{2}]$, so that
$$
\beta_1 \leq \beta_2 \quad \Longrightarrow\quad \tilde{f}_{\beta_1}
\geq \tilde{f}_{\beta_2} \quad \Longrightarrow\quad F(\beta_1) \geq
F(\beta_2).
$$
\end{proof}

\begin{remark}
The above corollary was the content of Theorem 5, the main  result in \cite{Kosaki:2005}
 and of Proposition IV.1  in
\cite{YanagiFuruichiKuriyama:2005}.  Note that, because of Corollary
\ref{biggest}, the optimal bound previously known was given by
$f_{WY}$, namely the bound of Wigner-Yanase metric (this was due to
Kosaki in \cite{Kosaki:2005}).  Remark \ref{sldwy} implies that the
bound given by the $SLD$ area is strictly greater then that given by
the $WY$ area.
\end{remark}

\begin{proposition} \label{dai}
$$\begin{array}{rcl}
{\rm Cov}_{\rho}(A,B)&=&{\rm Corr}_{\rho}^{f}(A,B)+{\cal C}^{\tilde
f}_{\rho}(A_0,B_0),\\[12pt] {\rm Var}_{\rho}(A)&=&I_{\rho}^f(A)+{\cal
C}^{\tilde f}_{\rho}(A_0). \end{array}$$
\end{proposition}
\begin{proof}
The calculations of Proposition \ref{nonso} imply that

\begin{align*}
{\rm Corr}_{\rho}^{f}(A,B)-{\rm Cov}_{\rho}(A,B)
&={\rm Tr}(\rho A ){\rm Tr}(\rho B)-{\rm Tr}(\Delta( A) B) \\
&=-{\rm Tr}(m_{\tilde f}(L_{\rho},R_{\rho})( A_0)B_0) \\
&=-{\cal C}^{\tilde f}_{\rho}(A_0,B_0).
\end{align*}
\end{proof}

Luo (see \cite{Luo:2005a}) suggested that if one consider the
variance as a measure of ``uncertainty" of an observable $A$ in the
state $\rho$ then the above equality splits the variance in a
``quantum" part ($I_{\rho}^f(A)$) plus a ``classical" part (${\cal
C}^{\tilde f}_{\rho}(A_0)$).

\section{Conditions for equality} \label{e}

In this section we give a necessary and sufficient condition to have
equality in our main result.

\begin{proposition}
    The inequality of Theorem \ref{main} is an equality if and only if
    $A_0$ and $B_0$ are proportional.
\end{proposition}
\begin{proof}
If $A_0= \lambda B_0$, with $\lambda\in \mathbb{R}$,  then

\begin{align*}
    {\rm Var}_{\rho} ( A ){\rm Var}_{\rho} ( B ) - |{\rm Re} \{ {\rm
    Cov}_{\rho} ( A,B)|^2 \} &={\rm Tr}(\rho A_0^2) {\rm Tr}(\rho
    B_0^2)-| {\rm Re} \{ {\rm Tr}(\rho A_0 B_0)\} |^2  \\
    &={\rm Tr}(\rho (\lambda B_0)^2) {\rm Tr}(\rho B_0^2)-|{\rm Re}
    \{{\rm Tr}(\rho \lambda B_0 B_0) \}|^2 \\
    &=\lambda^2 {\rm Tr}(\rho B_0^2)^2 -\lambda^2 |{\rm Tr}(\rho B_0^2
    )|^2 \\
    &=0.
\end{align*}

In this case the inequality is just the equality $0=0$.

Now we suppose that $A_0,B_0$ are not proportional and we prove that
the inequality is strict.  We use the same notations as in the proof
of Theorem \ref{main}.

Note that

$$
{\rm Var}_\rho \left( A \right){\rm Var}_\rho \left( B \right) -
\left| {{\mathop{\rm Re}\nolimits} \left\{ {{\rm Cov}_\rho \left(
{A,B} \right)} \right\}} \right|^2 - I_{\rho}^{f} \left( A
\right)I_{\rho}^{f} \left( B \right) + \left| {{\mathop{\rm
Re}\nolimits} \left\{ {{\rm Corr}_{\rho}^{f} \left( {A,B} \right)}
\right\}} \right|^2 =
$$
$$
=\xi - \eta =\frac{1}{4}\sum_{i,j,k,l}H_{
f}(\lambda_i,\lambda_j,\lambda_k,\lambda_l) \cdot K_{i,j,k,l}(A,B),
$$
and
$$
H_{f}(\lambda_i,\lambda_j,\lambda_k,\lambda_l) >0, \qquad \qquad
K_{i,j,k,l}(A,B) \geq 0 \qquad \qquad \forall i,j,k,l.
$$
Therefore, the strict inequality is equivalent to $\xi - \eta>0$,
which is, in turn, equivalent to
$$
K_{i,j,k,l}(A,B) > 0
$$
for some $i,j,k,l$.

From the fact that $A_0,B_0$ are not proportional one can derive that also the matrices $\{a_{ij}\}, \{b_{ij}\}$ are not proportional and
this implies (the other cases
being trivial) that there exist (complex) $a_{ij}, b_{ij}, a_{kl},
b_{kl} \not=0$ and (real) $\lambda, \mu \not=0$ such that
$$
a_{ij}=\lambda b_{ij} \qquad \qquad a_{kl}=\mu b_{kl} \qquad \qquad
\lambda\not= \mu.
$$
We get
\begin{align*}
    K_{i,j,k,l}(A,B) &=  |a_{ij} |^2 |b_{kl}|^2 + |a_{kl} |^2
    |b_{ij}|^2 - 2{\rm Re} \{ a_{ij} b_{ji} \} {\rm Re} \{ a_{kl}
    b_{lk} \}  \\
    &=  |a_{ij} |^2 |b_{kl}|^2 + |\mu b_{kl} |^2
    \big|\frac{a_{ij}}{\lambda}\big|^2- 2{\rm Re} \big\{ a_{ij}
    \frac{\overline{a_{ij}}}{\lambda} \big\} {\rm Re} \{ \mu b_{kl} b_{lk}
    \} \\
    &=\left(1+\frac{\mu^2}{\lambda^2}\right) \cdot |a_{ij} |^2
    |b_{kl}|^2 -2\frac{\mu}{\lambda} |a_{ij} |^2 |b_{kl}|^2 \\
    &=\left(1+\frac{\mu^2}{\lambda^2}-2\frac{\mu}{\lambda}\right)
    \cdot |a_{ij} |^2 |b_{kl}|^2 \\
    &=\left(1-\frac{\mu}{\lambda}\right)^2\cdot |a_{ij} |^2 |b_{kl}|^2
    >0 \\
\end{align*}
because
$$
\left(1-\frac{\mu}{\lambda}\right) \not= 0.
$$
Therefore,
$$
\xi - \eta \not= 0
$$
and this ends the proof.
\end{proof}

The particular case $f=f_{\beta}$ (where $\beta \in (0,1/2]$) of the
above proposition has been proved in Proposition 6 in
\cite{Kosaki:2005}.

\section{Another inequality} \label{another}

The study of the mean $m_{\tilde f}$ allows us to get another
inequality that can be seen as an uncertainty principle in Heisenberg
form. Recall that
$$
f_{RLD}(x):=\frac{2x}{x+1}.
$$

\begin{proposition}
$$
\Var_{\rho}(A) \geq I_{\rho}^f (A)+{\cal C}^{f_{RLD}}_{\rho}(A_0) \qquad \qquad \forall f \in {\cal F}_{op}.
$$
\end{proposition}
\begin{proof}
We use the notations employed in the proof of Theorem \ref{main}.  Since
$$\begin{array}{rcl}
\Var_{\rho}(A) &=& {\rm Tr} ( \rho A_0^2) =
\dfrac{1}{2}\displaystyle\sum_{i,j} ( \lambda _i + \lambda_j )
a_{ij} a_{ji}
\\[12pt]
I_{\rho}^f (A) &= &\Var_{\rho} ( A ) - {\rm Tr} ( A_0 m_{\tilde
f}(L_{\rho}, R_{\rho}) A_0) = \dfrac{1}{2}\displaystyle\sum_{i,j} (
\lambda _i + \lambda_j ) a_{ij} a_{ji} -\sum_{i,j} m_{\tilde
f}(\lambda_i,\lambda_j) a_{ij}a_{ji}
\\[12pt]
{\cal C}^{f_{RLD}}_{\rho}(A_0)&=&\displaystyle\sum_{i,j}
m_{h_0}(\lambda_i,\lambda_j) a_{ij}a_{ji}, \end{array}$$ using
Corollary \ref{max} we have
$$
\Var_{\rho}(A) - I_{\rho}^f (A)-{\cal C}^{f_{RLD}}_{\rho}(A_0) = \sum_{i,j}
[m_{\tilde f}(\lambda_i,\lambda_j)-m_{h_0}(\lambda_i,\lambda_j)] |a_{ij}|^2 \geq 0.
$$
\end{proof}

From this we get the following inequality.

\begin{theorem}
$$
\Var_{\rho}(A) \cdot \Var_{\rho}(B) \geq [I_{\rho}^f
(A)+{\cal C}^{f_{RLD}}_{\rho}(A_0)] \cdot [I_{\rho}^f (B)+{\cal C}^{f_{RLD}}_{\rho}(B_0)] \qquad \qquad \forall f \in {\cal F}_{op}.
\eqno{(9.1)}
$$
\end{theorem}

Since ${\cal C}^{f_{RLD}}_{\rho}(A_0) \geq 0$ we obtain, as a corollary, two
results due to Luo, for the case $f=f_{WY}=\frac14(1+\sqrt{x})^2$,
and to Hansen, for the general case (see \cite{Luo:2003},
\cite{Hansen:2006b}).

\begin{proposition}  \label{Luo:9.3}

$$
\Var_{\rho}(A) \geq I_{\rho}^f (A) \qquad \qquad \forall f \in {\cal F}_{op}.
$$
\end{proposition}

\begin{theorem}
\label{Hansen}

$$
\Var_{\rho}(A) \cdot \Var_{\rho}(B) \geq I_{\rho}^f (A) \cdot
I_{\rho}^f (B) = \frac{f(0)^2}{4} \cdot ||i[\rho,A]||^2_{\rho, f}
\cdot ||i[\rho,B]||^2_{\rho, f} \qquad \qquad \forall f \in {\cal F}_{op}.
$$
\end{theorem}

Let us study how the bound $I_{\rho}^f (A) \cdot I_{\rho}^f (B)$
depends on $f$.

\begin{proposition}
For any $f, g \in {\cal F}_{op}$
$$\begin{array}{rcl}
f \leq g \quad &\Longrightarrow &{\cal C}^{f}_{\rho}(A_0) \leq {\cal
C}^{g}_{\rho}(A_0) ,
\\[12pt]
\tilde{f} \leq \tilde{g} \quad &\Longrightarrow &I_{\rho}^f (A) \geq
I_{\rho}^g (A). \end{array}$$
\end{proposition}

\begin{proof}
We still use notations of Theorem \ref{main}. Since $m_{f} \leq m_{
g}$,

\begin{align*}
{\cal C}^{g}_{\rho}(A_0)-{\cal C}^{f}_{\rho}(A_0)
&=\sum_{i,j} m_{ g}(\lambda_i,\lambda_j) a_{ij}a_{ji}-\sum_{i,j} m_{
    f}(\lambda_i,\lambda_j) a_{ij}a_{ji} \\
&=\sum_{i,j} [m_{ g}(\lambda_i,\lambda_j) -m_{f}(\lambda_i,\lambda_j)]|a_{ij}|^2 \geq 0.
\end{align*}

The second inequality is an immediate consequence of the first one.
\end{proof}

\begin{corollary}
$$
I_{\rho}^{SLD} (A) \geq I_{\rho}^f (A) \qquad \qquad \forall f \in
{\cal F}_{op}.
$$
\end{corollary}
\begin{proof}
Immediate consequence of Proposition \ref{max}.
\end{proof}

\begin{corollary}
$$
\tilde{f} \leq \tilde{g} \quad \Longrightarrow I_{\rho}^f
(A)I_{\rho}^f (B) \geq I_{\rho}^g (A)I_{\rho}^g (B).
$$
\end{corollary}

We discuss, now, the equality in Theorem \ref{Hansen}.

\begin{proposition}
$$
\Var_{\rho}(A) \cdot \Var_{\rho}(B) = I_{\rho}^f (A) \cdot
I_{\rho}^f (B)\qquad \Longleftrightarrow\qquad A_0=B_0=0.
$$
\end{proposition}
\begin{proof}
Because of Proposition \ref{Luo:9.3} we have
$$
\Var_{\rho}(A) \cdot \Var_{\rho}(B) = I_{\rho}^f (A) \cdot I_{\rho}^f
(B)
\Longleftrightarrow
\Var_{\rho}(A)= I_{\rho}^f (A),\ \Var_{\rho}(B) =
I_{\rho}^f (B).
$$
Hence, we need to show $\Var_{\rho}(A)= I_{\rho}^f (A)
\Longleftrightarrow A_{0}=0$.  Indeed, using the same notations as
in Theorem \ref{main},
\begin{align*}
    \Var_{\rho}(A) = I_{\rho}^f (A) & \Longleftrightarrow {\rm Tr} (
    A_0 m_{\tilde f}(L_{\rho}, R_{\rho}) A_0)=0 \Longleftrightarrow
    \sum_{i,j} m_{\tilde f}(\lambda_i,\lambda_j) a_{ij}a_{ji} =0 \\
    & \Longleftrightarrow a_{ij} =0, \ \forall i,j \Longleftrightarrow
    A_0=0.
\end{align*}
\end{proof}

\section{Relation with the standard uncertainty principles} \label{counter}

Some authors tried to prove the following inequalities
$$
\left( \frac{f(0)}{2} \cdot {\rm Area}(i[\rho,A], i[\rho,B])\right)^2
= I_{\rho}^{f} \left( A \right)I_{\rho}^{f} \left( B \right) - |{\rm
Re}({\rm Corr}_{\rho}^f(A,B))|^2 \geq \frac{1}{4}\vert {\rm
Tr}(\rho[A,B])\vert^2, \eqno{(10.1)}
$$
$$
I_{\rho}^{f} \left( A \right)I_{\rho}^{f} \left( B \right) \geq
\frac{1}{4}\vert {\rm Tr}(\rho[A,B])\vert^2.  \eqno{(10.2)}
$$
They wanted to obtain the standard Heisenberg-Schr\"odinger
uncertainty principles as consequences of the uncertainty principles
discussed in the present paper.  Actually the inequality (10.1) has
been proved false for $f=f_{\beta}$, that is, for the
Wigner-Yanase-Dyson case (see p.632, 642-644 in \cite{Kosaki:2005},
p.4404 in \cite{YanagiFuruichiKuriyama:2005} and
\cite{LuoQZhang:2005}).  But the discussion of Section \ref{m},
\ref{corr}, \ref{another} shows that the upper bounds
$$
G(f)=\frac{f(0)}{2} \cdot {\rm Area}(i[\rho,A], i[\rho,B]) \qquad
\qquad N(f):=I_{\rho}^{f} \left( A \right)I_{\rho}^{f} \left( B
\right)
$$
can be larger than those of the $WYD$ metric (we showed it for the
$SLD$ metric in Remark \ref{sldwy}). It is, therefore, natural to
ask if the above inequalities, that are false for the $WYD$ metric,
can be true for some different quantum Fisher information (for
example for the $SLD$ metric).  The following theorem shows that
this is not the case, even on $2 \times 2$ matrices.

\begin{theorem} \label{contro}

    There exist $2 \times 2$ self-adjoint matrices $A$ and $B$, and a
    density matrix $\rho$ such that
$$
I_{\rho}^{f} \left( A \right)I_{\rho}^{f} \left( B \right) <
\frac{1}{4}\vert {\rm Tr}(\rho[A,B])\vert^2 \qquad \qquad \forall f
\in {\cal F}_{op}.
$$
Therefore, for these $\rho,A,B$ we also have
$$
\left( \frac{f(0)}{2} \cdot {\rm Area}(i[\rho,A], i[\rho,B])\right)^2
= I_{\rho}^{f} \left( A \right)I_{\rho}^{f} \left( B \right) - |{\rm
Re}({\rm Corr}_{\rho}^f(A,B))|^2 < \frac{1}{4}\vert {\rm
Tr}(\rho[A,B])\vert^2 \qquad \qquad \forall f \in {\cal F}_{op}.
$$
\end{theorem}
\begin{proof}
    We use notations of Theorem \ref{main}: let
    $\left\{\varphi_i\right\}$ be a complete orthonormal base composed
    of eigenvectors of $\rho$, and $\{ {\lambda}_i \}$ the
    corresponding eigenvalues.  Set $a_{ij} \equiv \langle {A_0}
    {\varphi}_i |{\varphi}_j \rangle $ and $ b_{ij} \equiv \langle B_0
    \varphi_i | {\varphi_j } \rangle $. In what follows $\lambda_1 > \lambda_2 >0$, $\lambda_1 + \lambda_2=1$
and
$$
\rho =\begin{pmatrix}
\lambda_1 & 0  \cr
0 & \lambda_2  \cr
\end{pmatrix},  \quad
A=\begin{pmatrix}
0 & i  \cr
-i & 0  \cr
\end{pmatrix}, \quad
B=\begin{pmatrix}
0 & 1 \cr
1 & 0  \cr
\end{pmatrix},
$$
(in terms of Pauli matrices, $A=-\sigma_2$ and $B=\sigma_1$). Simple
calculations show  that $|a_{ii}|=|b_{ii}|=0$, while
$|a_{ij}|=|b_{ij}|=1$ for any $i,j$ such that $i \not= j$.
Therefore,

\begin{align*}
{\rm Var}_{\rho}(A)
&={\rm Tr}(\rho A_0) \\
&=\frac{1}{2}\sum_{i,j} ( \lambda _i + \lambda_j )a_{ij} a_{ji} \\
&=\frac{1}{2}((\lambda_1 + \lambda_2)+(\lambda_2 +\lambda_1)) \\
&=1,
\\[18pt]
{\cal C}^{\tilde f}_{\rho}(A_0)
&=\sum_{i,j} m_{\tilde f}(\lambda_i,\lambda_j)a_{ij}a_{ji} \\
&= (m_{\tilde f}(\lambda_1,\lambda_2)+m_{\tilde f}(\lambda_1,\lambda_2)) \\
&= 2m_{\tilde f}(\lambda_1,\lambda_2),
\\[18pt]
I_{\rho}^f (A) &={\rm Var}_{\rho}(A)-{\cal C}^{\tilde f}_{\rho}(A_0)=1-2m_{\tilde f}(\lambda_1,\lambda_2).
\end{align*}

By the same reasoning,
\begin{align*}
{\rm Var}_{\rho}(B)&=1\\
{\cal C}^{\tilde f}_{\rho}(B_0)&=2m_{\tilde f}(\lambda_1,\lambda_2) \\
I_{\rho}^f (B) &=1-2m_{\tilde f}(\lambda_1,\lambda_2).
\end{align*}

Moreover, by direct calculation, one has that
$$
\frac{1}{4}\vert {\rm Tr}(\rho[A,B])\vert^2 =(\lambda_1-\lambda_2)^2.
$$

Now, recall that, since $m_{\tilde f}$ is a mean (and because of
Corollary \ref{basic}) one has for any $f \in {\cal F}_{op}$
$$
\lambda_1 > m_{\tilde f}(\lambda_1,\lambda_2) >\lambda_2 >0,
$$
$$
1-2m_{\tilde f}(\lambda_1,\lambda_2)=(\lambda_1 + \lambda_2)-2m_{\tilde f}(\lambda_1,\lambda_2) \geq 0.
$$
Hence, the following inequalities are equivalent
\begin{align*}
    I_{\rho}^{f} \left( A \right)I_{\rho}^{f} \left( B \right) & <
    \frac{1}{4}\vert {\rm Tr}(\rho[A,B])\vert^2 \\
    (1-2m_{\tilde f}(\lambda_1,\lambda_2))^2 &<
    (\lambda_1-\lambda_2)^2 \\
    (\lambda_1 + \lambda_2)-2m_{\tilde f}(\lambda_1,\lambda_2) & <
    \lambda_1-\lambda_2 \\
    2\lambda_2 & < 2m_{\tilde f}(\lambda_1,\lambda_2) \\
    \lambda_2 & < m_{\tilde f}(\lambda_1,\lambda_2),
\end{align*}
and so we get the conclusion.
\end{proof}

Note that
$$
\frac{1}{4}\vert {\rm Tr}(\rho[A,B])\vert^2 \geq I_{\rho}^{f} \left( A
\right)I_{\rho}^{f}\left( B \right)
$$
is obviously false, in general: if one takes $A=B$, the left side is
zero and the right side could be positive at the same time.

A similar argument applies to the inequality
$$
\frac{1}{4}\vert {\rm Tr}(\rho[A,B])\vert^2 \geq I_{\rho}^{f} \left( A
\right)I_{\rho}^{f}\left( B \right) - \left| {{\mathop{\rm
Re}\nolimits} \left\{ {{\rm Corr}_{\rho}^{f} \left( {A,B} \right)}
\right\}} \right|^2=\left(\frac{f(0)}{2}{\rm Area}_f
(i[\rho,A],i[\rho,B]) \right)^2;
$$
indeed, one may choose $\rho, A, B$ such that $[A,B]=0$ while
$[\rho,A],[\rho,B]$ are not proportional, so that they span a
positive area.

We may conclude that the Heisenberg and Schr\"odinger uncertainty
principles
$$\begin{array}{rcl}
{\rm Var}_\rho \left( A \right){\rm Var}_\rho \left( B \right)
&\geq& \dfrac{1}{4}\vert {\rm Tr}(\rho[A,B])\vert^2,
\\[12pt]
{\rm Area}^{{\rm Cov}^s}_{\rho}(A,B) &\geq &\dfrac{1}{2}\vert {\rm
Tr}(\rho[A,B])\vert, \end{array}$$ cannot be deduced from the
uncertainty principles
$$\begin{array}{rcl}
{\rm Var}_\rho \left( A \right){\rm Var}_\rho \left( B \right)
&\geq& I_{\rho}^{f} \left( A \right) \cdot I_{\rho}^{f}\left( B
\right),
\\[12pt]
{\rm Area}^{{\rm Cov}^s}_{\rho}(A,B) &\geq &\dfrac{f(0)}{2}\cdot
{\rm Area}^f_{\rho}(i[\rho,A],i[\rho,B]), \end{array}$$ and
vice versa.

\bigskip

The above described mistake appeared several times in the literature
(see Theorem 2 in \cite{Luo:2003}, Theorem 2 in
\cite{LuoZZhang:2004}, Theorem 1 in \cite{LuoQZhang:2004} and
Note 1, Section 3.2 in \cite{Hansen:2006b}).  It can be helpful to
explain its origin, again along the lines of \cite{Kosaki:2005} (see
also \cite{YanagiFuruichiKuriyama:2005}).

We have seen that
$$
\frac{1}{2i}{\rm Tr}(\rho[A,B])=\frac{1}{2i}({\rm
Corr}_{\rho}^f(A,B)-{\rm Corr}_{\rho}^f(B,A))= {\rm Im}({\rm
Corr}_{\rho}^f(A,B))
$$
and therefore
$$
\frac{1}{4}|{\rm Tr}(\rho[A,B])|^2=| {\rm Im}({\rm
Corr}_{\rho}^f(A,B))|^2 \leq |{\rm Corr}_{\rho}^f(A,B)|^2.
$$
If there were a Cauchy-Schwartz type estimate
$$
|{\rm Corr}_{\rho}^f(A,B)|^2 \leq {\rm Corr}_{\rho}^f(A,A) \cdot {\rm
Corr}_{\rho}^f(B,B) \eqno{(11.1)}
$$
using, for example, Theorem \ref{Hansen} one would get a refined
Heisenberg uncertainty principle in the form
$$
{\rm Var}_{\rho}(A) \cdot {\rm Var}_{\rho}(B) \geq I_{\rho}^{f} \left(
A \right)\cdot I_{\rho}^{f} \left( B \right) \geq \frac{1}{4}\vert
{\rm Tr}(\rho[A,B])\vert^2.
$$
By Theorem \ref{contro} we know that this is impossible.  The wrong
point is the Cauchy-Schwartz estimate (11.1), which is false.  This
depends on the following facts.  The sesquilinear form
$$
{\rm Corr}^f_{\rho}(X,Y):={\rm Tr}(\rho X^{\dag}Y)-{\rm Tr}(X^{\dag}
\cdot m_{\tilde f}(L_{\rho},R_{\rho})(Y))
$$
on the complex space $M_n$ is not positive (see p.  632 in
\cite{Kosaki:2005}).  On the other hand, $ {\rm Corr}^f_{\rho}(A,B)$
is not a real form on the real space $M_{n,sa}$: also in this case
one cannot prove the desired Cauchy-Schwartz inequality. The best
one can have is a Cauchy-Schwartz estimate only for the (real)
positive bilinear form $ {\rm Re}\{{\rm Corr}^f_{\rho}(A,B)\}$ on
$M_{n,sa}$ (see p.643 in \cite{Kosaki:2005} and
\cite{LuoQZhang:2005}).  This would imply simply
$$
\left( \frac{f(0)}{2 } \cdot {\rm
Area}^{f}_{\rho}(i[\rho,A],i[\rho,B]) \right)^2= I_{\rho}^{f} \left( A
\right)I_{\rho}^{f} \left( B \right) - \left| {{\mathop{\rm
Re}\nolimits} \left\{ {{\rm Corr}_{\rho}^{f} \left( {A,B} \right)}
\right\}} \right|^2 \geq 0.
$$

\section{Not faithful states and pure states} \label{p}

We discuss, now, the general case $\rho \geq 0$.

\begin{proposition}
    The function $m_{\tilde
    f}:(0,\infty)\times(0,\infty)\to(0,\infty)$ has a continuous
    extension to $[0,\infty)\times[0,\infty)$.
\end{proposition}
\begin{proof}
    If $f$ is regular then, for example,
$$
\lim_{(x,y) \to (0,y_0)} m_{\tilde f}(x,y)= \frac{y_{0}}{2} -
\frac{f(0)y_{0}^{2}}{2y_{0}f(0)} = 0.
$$

If $f$ is not regular then $m_{\tilde f}(x,y)=\frac{x+y}{2}$ and we
are done (see \cite{Hansen:2006b}).
\end{proof}

The definition of $f$-correlation still makes sense and the inequality
of Theorem \ref{main}
$$
{\rm Var}_\rho \left( A \right){\rm Var}_\rho \left( B \right) -
\left| {{\mathop{\rm Re}\nolimits} \left\{ {{\rm Cov}_\rho \left(
{A,B} \right)} \right\}} \right|^2 \geq I_{\rho}^{f} \left( A
\right)I_{\rho}^{f} \left( B \right) - \left| {{\mathop{\rm
Re}\nolimits} \left\{ {{\rm Corr}_{\rho}^{f} \left( {A,B} \right)}
\right\}} \right|^2
$$
holds by continuity for arbitrary (not necessarily faithful) states.

In what follows we study the pure state case.

\begin{corollary} \label{spectral}
If $s: [0,+\infty) \times [0,+\infty) \to \mathbb{R}$ is a continuous
function, and $\rho$ is a pure state, then
$$
s(L_{\rho},R_{\rho})(A) =\rho A \rho.
$$
\end{corollary}
\begin{proof}
Consequence of Corollary \ref{coro:spectral}.
\end{proof}

\begin{lemma}
If $\rho$ is pure, then ${\rm Tr}((\rho A \rho)(\rho B \rho))={\rm
Tr}(\rho A \rho)\cdot {\rm Tr}(\rho B \rho)$.
\end{lemma}
\begin{proof}
Suppose for simplicity that $\rho={\rm diag}(1,0,...,0)$ (the general
case follows easily from this).  Then $\rho A \rho= {\rm
diag}(A_{11},0,...,0)$ and the same holds for $B$.  Therefore $(\rho A
\rho )(\rho B \rho)= {\rm diag}(A_{11}B_{11},0,...,0)$.  This implies
$$
{\rm Tr}((\rho A \rho )(\rho B \rho))=A_{11}B_{11} ={\rm Tr}(\rho A
\rho ) \cdot {\rm Tr}(\rho B \rho).
$$
\end{proof}

\begin{lemma}
    If $\rho$ is pure, then
$$
{\rm Tr}(m_{f}(L_{\rho}, R_{\rho})(A) B)={\rm Tr}(\rho A) \cdot
{\rm Tr}(\rho B).
$$
\end{lemma}
\begin{proof}
By Corollary \ref{spectral} one has
$$
m_{ f}(L_{\rho}, R_{\rho})(A)=\rho A \rho
$$
and therefore
\begin{align*}
    {\rm Tr}(m_{f}(L_{\rho}, R_{\rho})(A) B) &={\rm Tr}(\rho A
    \rho B) \\
    &={\rm Tr}((\rho A \rho )(\rho B \rho)) \\
    &={\rm Tr}(\rho A \rho ) \cdot {\rm Tr}(\rho B \rho) \\
    &={\rm Tr}(\rho A ) \cdot {\rm Tr}(\rho B ).
\end{align*}
\end{proof}
\begin{corollary}
If $\rho$ is pure, then
$$
{\cal C}^f_{\rho}(A_0,B_0)={\rm Tr}(m_{f}(L_{\rho}, R_{\rho})(A_0) B_0)={\rm Tr}(\rho A_0) \cdot
{\rm Tr}(\rho B_0)=0.
$$
\end{corollary}

\begin{proposition}
If $\rho$ is pure, then
$$
{\rm Corr}^f_{\rho}(A,B)= {\rm Cov}_{\rho}(A,B) \qquad \qquad \forall
f \in {\cal F}_{op}.
$$
\end{proposition}
\begin{proof}
Immediate from the above Corollary and Proposition \ref{dai}

\end{proof}

The case $I^f_{\rho}(A)= {\rm Var}_{\rho}(A)$ was proved by Hansen in
Theorem 3.8 p.16 in \cite{Hansen:2006b}.

Therefore, on pure states we have the equalities
\[
\begin{array}{rcl}
{\rm Var}_\rho \left( A \right){\rm Var}_\rho \left( B \right) -
\left| {{\mathop{\rm Re}\nolimits} \left\{ {{\rm Cov}_\rho \left(
{A,B} \right)} \right\}} \right|^2 &=& I_{\rho}^{f} \left( A
\right)I_{\rho}^{f} \left( B \right) - \left| {{\mathop{\rm
Re}\nolimits} \left\{ {{\rm Corr}_{\rho}^{f} \left( {A,B} \right)}
\right\}} \right|^2,
\\[12pt]
{\rm Var}_\rho \left( A \right){\rm Var}_\rho \left( B \right) &=&
I_{\rho}^{f} \left( A \right)I_{\rho}^{f} \left( B \right) .
\end{array}
\] This implies that, if a sequence of faithful states $D_n$ converges
to the pure state $\rho$, then the limit

\begin{align*}
    \lim_{n \to +\infty} \left( \frac{f(0)}{2} \cdot {\rm
    Area}_{D_n}^f ( i[D_n,A], i[D_n, B] ) \right)^2 &= \lim_{n \to
    +\infty} I_{D_n}^{f} \left( A \right)I_{D_n}^{f} \left( B \right)
    - \left| {{\mathop{\rm Re}\nolimits} \left\{ {{\rm Corr}_{D_n}^{f}
    \left( {A,B} \right)} \right\}} \right|^2 \\
    &= I_{\rho}^{f} \left( A \right)I_{\rho}^{f} \left( B \right) -
    \left| {{\mathop{\rm Re}\nolimits} \left\{ {{\rm Corr}_{\rho}^{f}
    \left( {A,B} \right)} \right\}} \right|^2 \\
    &= {\rm Var}_\rho \left( A \right){\rm Var}_\rho \left( B \right)
    - \left| {{\mathop{\rm Re}\nolimits} \left\{ {{\rm Cov}_\rho
    \left( {A,B} \right)} \right\}} \right|^2
\end{align*}
does not depend on $f$.

This result has an interesting alternative explanation, using a
theorem by Petz and Sudar that describes the possible extension of
quantum Fisher information to pure states (see \cite{PetzSudar:1996}).
We devote the rest of the section to explain this phenomenon.

Let $M^0_n=M^0_n({\mathbb C})$ be the set of faithful states whose
eigenvalues are all distinct.  Recall that the pure states are
identified with ${\mathbb C}P^{n-1}$, the complex projective space.
On ${\mathbb C}P^{n-1}$ one has a natural metric, the Fubini-Study
metric (denoted by $\langle \cdot, \cdot \rangle_{\rho, FS}$).  We
denote by $D$ the elements of $M_n^0$ and by $\rho$ the elements of
${\mathbb C}P^{n-1}$.  We can define a projection $\pi:M^0_n \to
{\mathbb C}P^{n-1}$ as follows: $\pi(D) \in {\mathbb C}P^{n-1}$ is
the pure state associated to the one-dimensional eigenspace
corresponding to the largest eigenvalue of $D \in M^0_n$.  With this
definition, $\pi:M^0_n \to {\mathbb C}P^{n-1}$ is a smooth fiber
bundle.  The structure group is $U(1) \times U(n-1)$ (where $U(k)$
is the group of $k \times k$ unitary matrices).  The fiber space is
$\pi^{-1}(e)$ where $e$ is the ray generated by the vector $(1,0,
...  ,0) \in {\mathbb C}^{n}$.  Now, fix a monotone metric $\langle
\cdot, \cdot \rangle_{D,f}$.  We denote by $T_{D} \pi$ the
differential of $\pi$ at $D$ and let $H_{D}$ be the orthogonal
complement of $\hbox{ker}(T_{D} \pi)$ with respect to $\langle
\cdot, \cdot \rangle_{D,f}$.  Since $T_{D} \pi$ is surjective, the
restriction of $T_{D} \pi$ gives a linear isomorphism between
$H_{D}$ and $T_{\pi(D)}{\mathbb C}P^{n-1}$. For any tangent vector
$A \in T_{\pi(D)}{\mathbb C}P^{n-1}$ there is a unique ``lift" $A_D
\in H_{D} \subset T_{D}(M^0_n)$ such that $(T_{D} \pi)(A_D)=A$.

\begin{definition} \cite{PetzSudar:1996}
    We say that the sequence $D_n \in M^0_n$ radially converges to
    $\rho \in {\mathbb C}P^{n-1}$ if $D_n \to \rho$ as density
    matrices in $M^n$ and $\pi(D_n)=\rho$, $\forall n\in\bn$.
\end{definition}

\begin{definition} \cite{PetzSudar:1996}
    A metric $k$ on ${\mathbb C}P^{n-1}$ is a radial extension of a
    metric $g$ on $M^0_n$ if for any sequence $D_n \in M^0_n$, radially
    convergent to a point $\rho \in {\mathbb C}P^{n-1}$, and for any
    tangent vectors $A,B \in T_{\rho}{\mathbb C}P^{n-1}$, one has
$$
\lim_{n \to +\infty} g(A_{D_n}, B_{D_n})=k(A,B).
$$
\end{definition}

\begin{theorem} \cite{PetzSudar:1996}

    A monotone metric admits a radial extension if and only if it is
    regular, namely iff $f(0)\not=0$.  In this case the associated
    extension is just a multiple of the Fubini-Study metric according
    to the formula
$$
\lim_{n \to +\infty} \langle A_{D_n}, B_{D_n}
\rangle_{D_n,f}=\frac{1}{2f(0)} \langle A,B \rangle_{\rho,FS}.
$$
\end{theorem}

\begin{lemma} \cite{PetzSudar:1999}

    With the above definition,
$$
\pi(D)=\rho \qquad \Longrightarrow \qquad [D,A]=([\rho, A])_D,
$$
namely, the lift of commutator is the commutator of the lift.
\end{lemma}

This implies the following result.

\begin{proposition}
    If $D_n \to \rho$ radially then
$$
\lim_{n \to +\infty} f(0) \cdot {\rm Area}_{D_n}^f ( i[D_n,A], i[D_n,
B] ) = \frac{1}{2} \cdot {\rm Area}_{\rho}^{FS} ( i[\rho,A], i[\rho,
B] ).
$$
\end{proposition}

Hence, we have obtained the limit behavior by a totally different
argument.

\section{Optimality of an improvement for Heisenberg uncertainty principle} \label{Park}

The following result has been proved by Park in \cite{Park:2005} and indipendently by Luo in \cite{Luo:2005b}.

\begin{theorem}
If $g_0(x)=\sqrt{x}$ then

$$
{\rm Var}_{\rho}(A)\cdot{\rm Var}_{\rho}(B)
\geq
 {\cal C}^{g_0}_{\rho}(A_0){\cal C}^{g_0}_{\rho}(B_0)+\frac{1}{4}\vert{\rm Tr}(\rho[A,B])\vert^2.
$$
\end{theorem}

Note that the term ${\cal C}^{g_0}_{\rho}(A_0){\cal C}^{g_0}_{\rho}(B_0)$ disappears for pure states.
We prove that the above result is the best one can have considering functions $f \in {\cal F}_{op}$.

\begin{theorem}

For any $f \in {\cal F}_{op}$ we have

$$
{\rm Var}_{\rho}(A)\cdot{\rm Var}_{\rho}(B)
\geq
 {\cal C}^{f}_{\rho}(A_0){\cal C}^f{}_{\rho}(B_0)+\frac{1}{4}\vert{\rm Tr}(\rho[A,B])\vert^2 \qquad \Longleftrightarrow \qquad f(x) \leq  \sqrt{x}.
$$
\end{theorem}
\begin{proof}
We have
$$
f(x) \leq  g(x)\quad \Longrightarrow \quad m_f(x,y)\leq m_g(x,y)\quad \Longrightarrow \quad {\cal C}^f_{\rho}(A_0) \leq {\cal C}^g_{\rho}(A_0)
$$
and therefore if $f(x) \leq  \sqrt{x}$ we are done.

If $f(x_0) >  \sqrt{x_0}$ for a certain $x_0$ we produce a
counterexample. To this end, we do the same we did in the proof of
Theorem \ref{counter}.

Consider again $\lambda_1 > \lambda_2 >0$, $\lambda_1 + \lambda_2=1$ and
$$
\rho =\begin{pmatrix}
\lambda_1 & 0  \cr
0 & \lambda_2  \cr
\end{pmatrix},  \quad
A=\begin{pmatrix}
0 & i  \cr
-i & 0  \cr
\end{pmatrix}, \quad
B=\begin{pmatrix}
0 & 1 \cr
1 & 0  \cr
\end{pmatrix}.
$$

We have calculated
$$
{\rm Var}_{\rho}(A)=1 \qquad \qquad {\cal C}^f_{\rho}(A_0)=2m_{ f}(\lambda_1,\lambda_2)
$$
$$
{\rm Var}_{\rho}(B)=1 \qquad \qquad {\cal C}^f_{\rho}(B_0)=2m_{
f}(\lambda_1,\lambda_2),
$$
$$
{\rm Tr}(\rho[A,B])=(\lambda_1-\lambda_2)^2.
$$

In this case the inequality
$$
{\rm Var}_{\rho}(A)\cdot{\rm Var}_{\rho}(B)
\geq
 {\cal C}^f_{\rho}(A_0){\cal C}^f_{\rho}(B_0)+\frac{1}{4}\vert{\rm Tr}(\rho[A,B])\vert^2
$$
reads as
$$
1 \cdot 1 \geq 2m_{ f}(\lambda_1,\lambda_2)\cdot 2m_{
f}(\lambda_1,\lambda_2)+(\lambda_1-\lambda_2)^2,
$$
that is,
$$
1 \geq 4(m_{ f}(\lambda_1,\lambda_2))^2+(\lambda_1-\lambda_2)^2
$$
or
$$
1 \geq 4 \left(\lambda_2 f \left(
\frac{\lambda_1}{\lambda_2}\right)\right)^2+(\lambda_1-\lambda_2)^2.
$$
For $g_0(x)=\sqrt{x}$ we have
$$
4 \left(\lambda_2 g_0 \left(
\frac{\lambda_1}{\lambda_2}\right)\right)^2+(\lambda_1-\lambda_2)^2=1.
$$
Therefore, if for some $x_0 \not= 1$ we have $f(x_0) > \sqrt{x_0}$
then for $\frac{\lambda_1}{\lambda_2}=x_0$
\begin{align*}
{\cal C}^f_{\rho}(A_0){\cal C}^f_{\rho}(B_0)+\frac{1}{4}\vert{\rm Tr}(\rho[A,B])\vert^2
&=4 \left(\lambda_2 f \left( \frac{\lambda_1}{\lambda_2}\right)\right)^2+(\lambda_1-\lambda_2)^2 \\
&> 4 \left(\lambda_2 \sqrt{ \left( \frac{\lambda_1}{\lambda_2}\right)}\right)^2+(\lambda_1-\lambda_2)^2 \\
&=1 \\
&={\rm Var}_{\rho}(A)\cdot{\rm Var}_{\rho}(B)
\end{align*}

that is, the inequality is false.

\end{proof}

\noindent
{\Large {\bf Acknowledgements}}

It is a pleasure to thank Frank Hansen for sending us the preprint \cite{Hansen:2006b}.

\end{document}